\documentclass[epj
]{svjour}
\usepackage{amsmath,amssymb}
\usepackage{graphicx}
\usepackage{dcolumn}
\usepackage{bm}
\PassOptionsToPackage{hyphens}{url}
\usepackage{hyperref}
\usepackage{xcolor}
\usepackage{lipsum}
\usepackage{subcaption}
\usepackage{cite}
\usepackage{cuted}
\begin{document}
\title{Normalizing Flow-Assisted Nested Sampling on Type-II Seesaw Model}
\author{Rajneil Baruah\inst{1} \and Subhadeep Mondal\inst{1} \and Sunando Kumar Patra\inst{2} \and Satyajit Roy\inst{2}
}                     
%
%
\institute{Department of Physics, SEAS, Bennett University, Greater Noida, Uttar Pradesh, 201310, India.
\and Department of Physics, Bangabasi Evening College, Kolkata, 700009, West
Bengal, India}
\date{Received: date / Revised version: date}
%
\abstract{
We propose a novel technique for sampling particle physics model parameter space. The main sampling method applied is Nested Sampling (NS), which is boosted by the application of multiple Machine Learning (ML) networks, e.g., Self-Normalizing Network (SNN) and Normalizing Flow (specifically RealNVP). We apply this to the Type-II Seesaw model to test the algorithm's efficacy. We present the results of our detailed Bayesian exploration of the model parameter space subjected to theoretical constraints and experimental data corresponding to the 125 GeV Higgs boson, $\rho$-parameter, and the oblique parameters.\\
All associated data, figures, and trained ML models can be found here:  \href{https://github.com/sunandopatra/MLNS-T2SS}{GitHub  
\includegraphics[height=\fontcharht\font`\B]{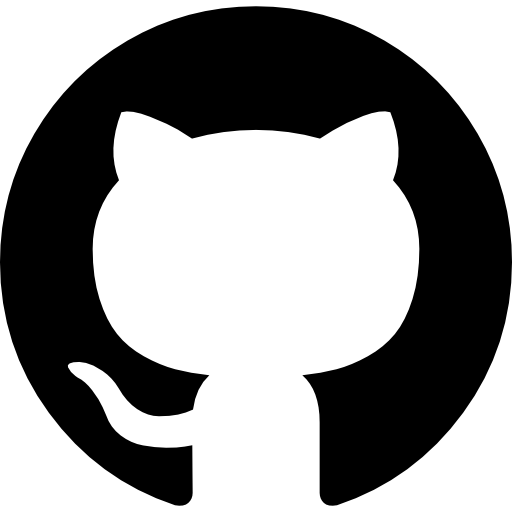}
}
\PACS{
      {PACS-key}{discribing text of that key}   \and
      {PACS-key}{discribing text of that key}
     } 
} 
\maketitle
%

\section{Introduction}\label{sec:intro}
Even as the large hadron collider (LHC) approaches its high luminosity era, a smoking gun signal for any beyond the standard model (BSM) physics scenario continues to elude us. However, we have a wealth of experimental data not only from the LHC but from other particle physics experiments as well, such as dark matter, neutrino oscillation, lepton flavour violating (LFV) and lepton number violating (LNV) decay search experiments, etc. As a result, phenomenological BSM searches have shifted more towards constraining new physics (NP) domains using the available data rather than predicting rates of new experimental observables by choosing random benchmark points from different regions of parameter space (PS). Even if one takes the second approach, one must have a clear idea about the available parameter space, consistent with existing data. This becomes increasingly difficult with the increasing dimensionality of NP parameter space. A large number of input parameters coupled with the growing precision of data makes the parameter space of most BSM scenarios intractable. 

The most common statistically viable way of handling this problem is a likelihood-based estimation (the likelihood is computed with the help of experimental data on a set of relevant observables), where one either finds the best-fit point in the PS with the associated confidence intervals by optimizing the negative log-likelihood or calculates the posterior distribution of the parameters. The most commonly used method to find the (unnormalized) posterior is Markov Chain Monte Carlo (MCMC), with all of its variants \cite{Hastings:1970,Speagle:2019ffr,brooks2011eds}, which, while not suitable to find the exact evidence in favour of the data, supplies a sample from the posterior. This enables one to plot one- or two-dimensional marginal credible interval contours from the posterior. These tell us which part of the parameter space one should concentrate upon for any phenomenological study. However, a well-known issue with MCMC is its tendency to get stuck in one mode of the PS, provided it finds one \cite{tripuraneni2017magnetic}. This becomes problematic for multi-modal spaces typical for many particle physics models \cite{Albert:2024zsh,GAMBIT:2017zdo,Kvellestad:2017rwl}. As a Markov chain, MCMC is, in theory, capable of finding the other modes, but takes an \textit{a priori} unknown amount of time in the process. A single chain is not parallelizable and methods involving multiple chains \cite{foreman2013emcee,DeBlas:2019ehy} increase the computational cost. Convergence criteria for these chains are neither foolproof nor beyond debate \cite{Cowles01081999}. The scenario worsens in the presence of typical theoretical constraints which prevent high-energy phenomenological (HEP) spectrum generators to generate consistent spectrum in a large chuck of the parameter space. All of these, in aggregate, make MCMC computationally expensive for complicated BSM parameter spaces. Application of machine learning (ML) algorithms to improve upon the existing techniques in this regard has gained relevance in recent times \cite{Caron:2019xkx,Hollingsworth:2021sii,DBLP:journals/corr/abs-2107-03342,Hammad:2022wpq,Goodsell:2022beo,Diaz:2024sxg,Binjonaid:2024jpm,Hammad:2024tzz,Hermans:2019ioj,Brehmer:2020cvb,Morrison:2022vqe,Hunt-Smith:2023ccp,Albergo:2021vyo}. For a more comprehensive account of these works one may refer to \cite{Baruah:2024gwy}. 

Newer, more efficient statistical techniques such as \textit{Nested Sampling} (NS) \cite{Skilling:2004,Skilling:2006} (auto-tuned convergence, access to multiple modes, parallelizability) are needed to circumvent the issues MCMC has. NS is a numerical integration strategy that can be used to obtain Bayesian \textit{evidence} and, in the process, one gets a weighted sample from the posterior. A well-known HEP package GAMBIT \cite{GAMBIT:2017yxo,Bloor:2021gtp} uses MultiNest \cite{Feroz:2008xx,Feroz:2009} to perform NS on BSM parameter space. However, as the algorithm moves toward high-likelihood regions, the progress becomes slower (and convergence takes longer) because of the rarity of higher likelihood points. We propose an ML solution to this. A special type of artificial neural network (ANN) is trained as a classifier to circumnavigate the external constraints, and a trained Normalizing Flow (NF)-type network is used to reduce the convergence time. The latter network is trained iteratively during the running of NS, such that, by the time NS reaches high-likelihood regions, it is trained well enough to predict a much larger number of higher likelihood points than a naive sampling of the prior. 

We have adopted a well-motivated BSM scenario, namely, the Type-II seesaw model to showcase the efficacy of this methodology. Type-II seesaw mechanism \cite{Konetschny:1977bn,Magg:1980ut,Schechter:1980gr,Lazarides:1980nt,Mohapatra:1980yp,Cheng:1980qt} provides us with one of the simplest frameworks to generate light neutrino masses and mixing angles \cite{RENO:2018dro,DayaBay:2018yms,Super-Kamiokande:2019gzr,T2K:2018rhz,T2K:2018rhz,NOvA:2019cyt,deSalas:2020pgw} without introducing any right-handed neutrinos into the theory. Instead, this model introduces a complex scalar, a triplet under the ${\rm SU(2)_L}$ gauge group. A Majorana mass term is generated for the light neutrinos after electroweak symmetry breaking through an LNV Yukawa coupling and the triplet vacuum expectation value (VEV). The model has very rich phenomenological implications \cite{Huitu:1996su,Chakrabarti:1998qy,Muhlleitner:2003me,Akeroyd:2005gt,Perez:2008ha0,delAguila:2008cj,Akeroyd:2009hb,Melfo:2011nx,Arhrib:2011vc,Akeroyd:2012nd,Chun:2012zu,Chun:2012jw,Englert:2013wga,Kanemura:2013vxa,Chun:2013vma,Chun:2003ej,Dey:2008jm,Arhrib:2011uy,Dev:2013ff,Arhrib:2014nya,Das:2016bir,Anisha:2021fzf,Du:2018eaw,Li:2018jns,Kanemura:2014goa,Han:2015hba,Han:2015sca,Mitra:2016wpr,Babu:2016rcr,BhupalDev:2018tox,Antusch:2018svb,Ferreira:2019qpf,Banerjee:2024jwn,Ashanujjaman:2023tlj,Giarnetti:2023dcr,Chiang:2021lsx,Ashanujjaman:2022ofg,Ashanujjaman:2022tdn,Ashanujjaman:2021txz,Ducu:2024xxf,Primulando:2019evb,Das:2023tna,Das:2024yvt,Das:2024kyk} owing to the exotic scalar sector and neutrino mass generation. The scalar sector consists of one doubly charged scalar, one singly charged scalar, and three neutral scalars: two CP-even and one CP-odd. In addition to that, lepton number-violating interactions lead to predictions of lepton number and lepton flavour-violating decays. All these factors combine to make Type-II seesaw a highly motivated and phenomenologically interesting scenario. The triplet VEV ($v_T$) is instrumental in this context. A sufficiently small $v_T$ ($\lesssim 10^{-9}$) allows the lepton number violating Yukawa couplings to be large enough in the context of light neutrino mass and mixing, but that also in turn leads to stringent constraints arising from non-observation of lepton number and/or lepton flavour violation. Such small $v_T$, however, does not affect electroweak precision data such as $W$ and $Z$ boson masses, $\rho$ parameter, etc. In this work, we explore the large $v_T$ scenario and ascertain how the parameter space shapes up when confronted with the electroweak precision data \cite{Amaldi:1987fu,Costa:1987qp,Langacker:1991an,Erler:1994fz,Haller:2018nnx,ParticleDataGroup:2022pth,ParticleDataGroup:2024cfk} and the 125 GeV Higgs data \cite{CMS:2012qbp,ATLAS:2012yve,ATLAS:2018tdk,CMS:2020xrn,ATLAS:2021vrm,cms_web1}. We also estimate the allowed numerical ranges of the LNV Yukawa parameters subjected to the neutrino oscillation data \cite{deSalas:2020pgw}. Although this scenario has been explored before, a posterior distribution with numerical estimates of central tendency and dispersion for all the new parameters introduced in this model is presented here for the first time.   

The article is organized as follows. Section~\ref{sec:theory} briefly describes the Type-II Seesaw model. Section~\ref{sec:method} starts with a discussion about the observable set and parameter ranges used in the analysis. Then we give a detailed account of the algorithm: the ML networks and their performance in particular. In section~\ref{sec:results}, after presenting the parameter posterior, we proceed to discuss the properties of the scalar particles in the model within the favoured parameter space and their collider constraints. We conclude in section~\ref{sec:concl}.

\section{Theory}\label{sec:theory} 
In addition to the ${\rm SU(2)_L}$ Higgs doublet ($\Phi$), Type-II Seesaw mechanism introduces an additional scalar triplet ($\Delta$) under the same gauge group with hypercharge $2$ (under the convention $Q=T_3+\frac{Y}{2}$). 
\begin{equation}
\Phi = \begin{pmatrix}
\phi^+  \\[8pt]
\phi^0
\end{pmatrix};    
\hspace{1cm}
\Delta= \begin{pmatrix}
\frac{1}{\sqrt{2}}\Delta^{+} & \Delta^{++} \\[8pt]
\Delta^0 & -\frac{1}{\sqrt{2}}\Delta^{+}
\end{pmatrix}  
\end{equation}
The scalar potential can be written as
\begin{align}
V(\Phi, & \Delta) \nonumber\\
&= -m^2_{\Phi} \Phi^{\dagger}\Phi + m^2_{\Delta}{\rm Tr}\Delta^{\dagger}\Delta + \left(\mu_1\Phi^Ti\sigma^2\Delta^{\dagger}\Phi + {\rm h.c}\right) \nonumber\\
&+ \frac{\lambda}{4} (\Phi^{\dagger}\Phi)^2 + \lambda_1\Phi^{\dagger}\Phi {\rm Tr}\Delta^{\dagger}\Delta + \lambda_2 ({\rm Tr}\Delta^{\dagger}\Delta)^2 \nonumber\\
&+ \lambda_3{\rm Tr}(\Delta^{\dagger}\Delta)^2 + \lambda_4\Phi^{\dagger}\Delta\Delta^{\dagger}\Phi\,.
\end{align}
Here $\sigma^2$ indicates the second Pauli matrix. $\mu_1$ is a trilinear coupling, whereas, $\lambda$ and $\lambda_i~(i=1,2,3,4)$ represent the quartic couplings. After electroweak symmetry breaking $\Phi$ and $\Delta$ acquire VEVs $\frac{v_d}{\sqrt{2}}$ and $\frac{v_T}{\sqrt{2}}$ respectively, with the total electroweak VEV given by $v=\sqrt{v_d^2 + 2v_T^2}$. The parameters $m^2_{\Phi}$ and $m^2_{\Delta}$ are not independent. They can be determined in terms of other free parameters using the tadpole equations:
\begin{eqnarray}
m^2_{\Phi} = \frac{\lambda_1 + \lambda_4}{2}v_T^2 - \sqrt{2}\mu_1v_T + \frac{\lambda}{4}v_d^2 \nonumber \\
m^2_{\Delta} = - \frac{\lambda_1 + \lambda_4}{2}v_d^2 + \frac{v_d^2\mu_1}{\sqrt{2}v_T} - (\lambda_2 + \lambda_3) v_T^2.
\end{eqnarray}

In the presence of the triplet VEV, at the tree level, the $W$ and $Z$ boson masses can be written as 
\begin{eqnarray}
m_W = \sqrt{\frac{g^2}{4} (v_d^2 + 2v_T^2)} \nonumber \\
m_Z = \sqrt{\frac{g^2+g^{\prime 2}}{4}(v_d^2 + 2v_T^2)}
\end{eqnarray}
where the $g$ and $g^{\prime}$ represent gauge couplings corresponding to ${\rm SU(2)_L}$ and ${ U(1)_Y}$ respectively.  The $\rho$-parameter can be written as
\begin{eqnarray}
\rho = 1 - \frac{2v_T^2}{v_d^2 + 4v_T^2}    
\end{eqnarray}
Unless $v_T << v_d$, the $\rho$-parameter is affected, and given that it is very precisely measured, one can restrict the triplet VEV from above effectively using this measurement. 

In the scalar sector, the neutral CP-even states can have considerably large mixing depending on the choices of new physics parameters. The mass-squared matrix at the tree level can be written as 
\begin{strip}
\rule[-1ex]{\columnwidth}{1pt}\rule[-1ex]{1pt}{1.5ex}
\begin{eqnarray}
m^2_{\rm CP-even} = \begin{pmatrix}
\frac{\lambda}{2}v_d^2 & -\sqrt{2}\mu_1 v_d + (\lambda_1 + \lambda_4)v_T v_d \\
-\sqrt{2}\mu_1 v_d + (\lambda_1 + \lambda_4)v_T v_d & \frac{\mu_1 v_d^2}{\sqrt{2}v_T} + 2(\lambda_2 + \lambda_3)v_T^2 
\end{pmatrix}    
\end{eqnarray}
\hfill\rule[1ex]{1pt}{1.5ex}\rule[2.3ex]{\columnwidth}{1pt}
\end{strip}
However, a large mixing between the CP-even states is disfavoured from the 125 GeV Higgs data. The mixing angle ($\alpha$) is given as 
\begin{eqnarray}
\tan 2\alpha = \frac{2(\lambda_1 + \lambda_4)v_Tv_d - 2\sqrt{2}v_d\mu_1}{\frac{\lambda v_d^2}{2}-\frac{\mu_1v_d^2}{\sqrt{2}v_T}-2(\lambda_2+\lambda_3)v_T^2}    
\label{eq:mix_ang}
\end{eqnarray}

The mixing angle between CP-odd states on the other hand is small unless $v_T\sim v_d$. The mass-squared matrix at the tree level can be written as 
 \begin{eqnarray}
m^2_{\rm CP-odd} = \begin{pmatrix}
2\sqrt{2}\mu_1 v_T & -\sqrt{2}\mu_1 v_d \\
-\sqrt{2}\mu_1 v_d  & \frac{\mu_1 v_d^2}{\sqrt{2}v_T} 
\end{pmatrix}    
\end{eqnarray}
Upon diagonalization, one obtains the only massive CP-odd scalar mass to be $m^2_A=\frac{\mu}{\sqrt{2}v_T}(v_d^2 + 4v_T^2)$. The charged scalar mass matrix can be written as 
\begin{align}
    m^2&_{\rm Charged-higgs} = \nonumber\\
    &\begin{pmatrix}
        \sqrt{2}\mu_1 v_T - \frac{\lambda_4}{2}v_T^2 & -\mu_1 v_d + \frac{\sqrt{2}}{4}\lambda_4v_T v_d \\
        -\mu_1 v_d + \frac{\sqrt{2}}{4}\lambda_4v_T v_d & \frac{\mu_1 v_d^2}{\sqrt{2}v_T} - \lambda_4v_d^2
    \end{pmatrix}    
\end{align}
with the charged Higgs mass 
\begin{eqnarray}
  m^2_{h^{\pm}}=\frac{2\sqrt{2}\mu_1 - \lambda_4v_T}{4v_T}(v_d^2+2v_T^2)\,.  
\label{eq:charged_mass}
\end{eqnarray}

The five quartic couplings introduced in the scalar potentials, namely, $\lambda$, $\lambda_1$, $\lambda_2$, $\lambda_3$ and $\lambda_4$ are strictly not independent of each other. Their choices are restricted theoretically from the requirements of vacuum stability and perturbative unitarity. These constraints have been studied in detail in the context of the Type-II seesaw model \cite{Bonilla:2015eha,Arhrib:2011uy}. For the constraint equations applicable to our choice of quartic coupling assignment, please refer to \cite{Primulando:2019evb}.  

The light neutrino masses are generated in this model through the interaction of the triplet scalar with the lepton doublet ($L$).
\begin{eqnarray}
L_{\rm neutrino}=Y_{ij}L^T_iCi\sigma^2\Delta L_j + {\rm h.c}    
\end{eqnarray}
Here $Y_{ij}$ is the a symmetric $3\times 3$ matrix with $i,j=1,2,3$. The Majorana mass terms for light neutrinos are generated after electroweak symmetry breaking. The mass matrix can be written as
\begin{eqnarray}
  M_{\nu} = \sqrt{2}Yv_T  
\label{eq:neut_mass}
\end{eqnarray}
Clearly, the only two parameters relevant for neutrino mass (and mixing angle) generation are the Yukawa couplings and the triplet VEV. 

\begin{figure*}[t]
    \centering
    \includegraphics[width=0.8\linewidth]{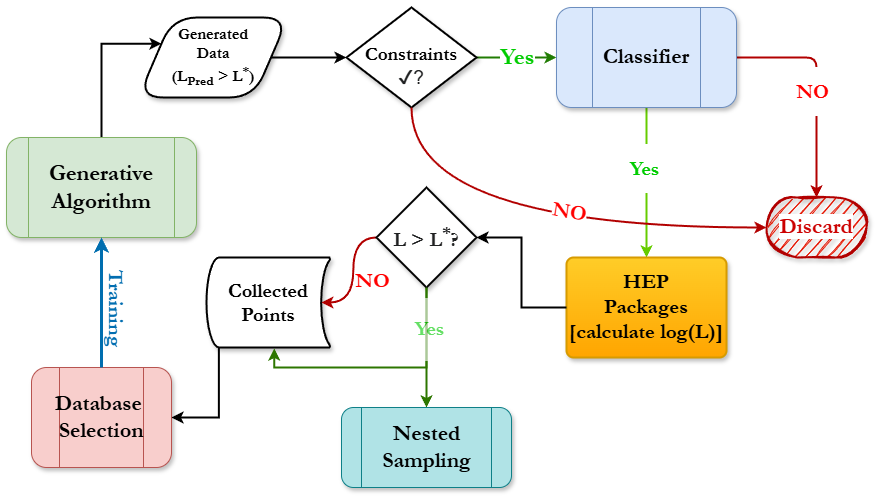}
    \caption{\small Detailed flow chart of our ML-assisted nested sampling algorithm to sample the parameter space.}
    \label{fig:flow1}
\end{figure*}

\section{Methodology}\label{sec:method} 
\subsection{Observables and Parameters}\label{sec:setup1}

The model is implemented in the \textit{Mathematica}\textsuperscript{\texttrademark} package \texttt{SARAH}\footnote{Version: \texttt{4.14.4}} \cite{Staub:2008uz,Staub:2010jh,Staub:2015kfa} which performs analytical calculations about mass matrices and vertices up to one loop. Full two-loop corrections are implemented for the scalar sector  \cite{Goodsell:2014bna,Goodsell:2015ira}. \texttt{SARAH} also generates source codes for \texttt{SPheno} \cite{Porod:2003um,Porod:2011nf} based on these analytical calculations. \texttt{SPheno} performs the numerical computation to generate a full particle spectrum including masses, mixing, and decay information of all particles. \texttt{SPheno} also provides the platform to check low-energy flavour constraints through its interface with \texttt{flavourKit} \cite{Porod:2014xia}. We have used the spectrum file generated by \texttt{SPheno}\footnote{Version: \texttt{4.0.4}} to check constraints from the SM-like Higgs data through \texttt{HiggsBounds} (HB) \cite{Bechtle:2008jh,Bechtle:2011sb,Bechtle:2012lvg,Bechtle:2013wla,Bechtle:2015pma,Bechtle:2020pkv}. HB determines whether a parameter point is excluded or not at $95\%$ confidence level (CL) based on the scalar search data obtained from LEP, Tevatron and the LHC. For likelihood calculation of the scalar sector at each parameter point we have used \texttt{HiggsSignals} (HS) \cite{Bechtle:2012lvg,Bechtle:2013xfa,Stal:2013hwa,Bechtle:2014ewa,Bechtle:2020uwn}.  
HS provides a \textit{peak-centered} $\chi^2$ measure of compatibility between the measured (SM-like) Higgs signal rates and masses obtained from the LHC and Tevatron and the model predictions for a fixed (hypothetical) Higgs mass. The combination of these two packages was used in our analysis as a \texttt{C++} implementation \texttt{HiggsTools}\footnote{Combination of \texttt{HiggsBounds-5} and \texttt{HiggsSignals-2}} \cite{Bahl:2022igd}. 

We have considered 163 observables in total: the full set of 159 Higgs data used by \texttt{HiggsSignals}, the three oblique parameters and the $\rho$-parameter. The Higgs sector observables are calculated by HS using the spectrum generated by \texttt{SPheno}. The rest of the observables are computed up to one-loop level by \texttt{SPheno}. 
The oblique parameter fits are performed by the \texttt{GFitter} collaboration \cite{Haller:2018nnx} and we use the data and correlation from there. The $\rho$ parameter data is obtained from \cite{ParticleDataGroup:2022pth,ParticleDataGroup:2024cfk}.

We have varied seven input parameters independently in this analysis. These are: $v_T$, $\lambda_1$, $\mu_1$, $\lambda_2$, $\lambda_3$, $\lambda_4$, and $\lambda$. We have kept the parameter domain as exhaustive as possible and ensured that each set of parameters chosen is consistent with vacuum stability and perturbative unitarity constraints. The parameter domain explored in the present work is: 

\begin{tabular}{lll}
     $v_T$ : (0.0001, 3.5), & $\lambda_1$ : (-3, 5), & $\mu_1$ : (-50, 200), \\
     $\lambda_2$ : (-3, 20), & $\lambda_3$ : (-25, 6), & $\lambda_4$ : (-5, 5), \\
     $\lambda$ : (0.01, 0.7). & &
\end{tabular}\\

While determining our explorable parameter space, we put a couple of loose constraints: the SM-like Higgs mass should lie within $90-160$ GeV and the $\rho$ parameter should be within $0.98-1.02$ ($\approx 100 \sigma$ on both sides of the mode).

\subsection{Framework}\label{sec:setup2} 
\subsubsection{Set-up}\label{sec:setup2a} 

We aim to perform an NS over the parameter space, to obtain a Bayesian posterior distribution dictated by the data used. Details about NS can be found in ref.s \cite{Skilling:2004,Skilling:2006,Ashton:2022,Baruah:2024gwy}. In short, the algorithm starts with $n_{Live}$ `\textit{live}' points sampled from a prior over the parameter space, generates more independent and identically distributed (\textit{iid}) points in a specified region until it finds one point with likelihood ($\mathcal{L}$) greater than the lowest likelihood among the live points ($\mathcal{L^*}$), replaces that lowest-likelihood point with the new one, saves the replaced point as a weighted sample, shrinks the target region in a specified way and then repeats. This way, evaluating a multi-dimensional integral essentially changes to a single-variable one, by peeling off equal-probability-contours from the parameter space. Mainly an algorithm to calculate the evidence/marginal likelihood of the posterior, NS generates a weighted sample drawn from the posterior, as a by-product. In addition to being self-tuning, mostly parallelizable, and able to sample from the whole prior space simultaneously---thus handling multi-modal problems much better than its MCMC counterparts---NS is also much faster than other equivalent MCMC methods while spending less computational resources. Due to the way NS works, it gets slower, the closer it gets to the high-likelihood region, especially if the shrinkage of the sampling region is comparatively less.

The main bottleneck in this kind of work is calculating the complicated likelihood function repetitively for different parameter points. Another additional source of slowdown is that not all points are allowed in the parameter space. Some of them are discarded due to directly applicable (thus quickly calculable) theoretical constraints; others, though allowed by the constraints mentioned, generate no spectrum (by some HEP spectrum generator like \texttt{SPheno}, mainly due to unstable vacuum and tachyonic masses in the spectrum). ML, especially generative algorithms, can provide speed-ups here. In addition to the application of a classifier to predict the valid points, the most obvious solution to learn the data-consistent parameter space is to learn the structure of the optimization criteria (negative log-likelihood or $\chi^2$ in our context) shared by similar points in the parameter space, known as `\textit{Amortized Optimization}' elsewhere in the literature \cite{amos2023tutorial}. The prediction uncertainty notwithstanding, this approach has numerous problems:
\begin{itemize}
    \item As it solely depends on the initial training data, the possibility of completely missing the high-likelihood region increases (especially if the likelihood is a highly varying function of the parameters). 
    \item The posteriors obtained are purely predictions. Checking them with reality then becomes the most time-consuming step in the algorithm.
    \item Predictions for degenerate parameter spaces are, at best, incomplete and at worst, incorrect, e.g., training on spaces with quadratic sign-ambiguity will learn neither of the modes, instead returning a region halfway between them.
\end{itemize}

\begin{figure*}
    \centering
    \includegraphics[width=0.45\linewidth]{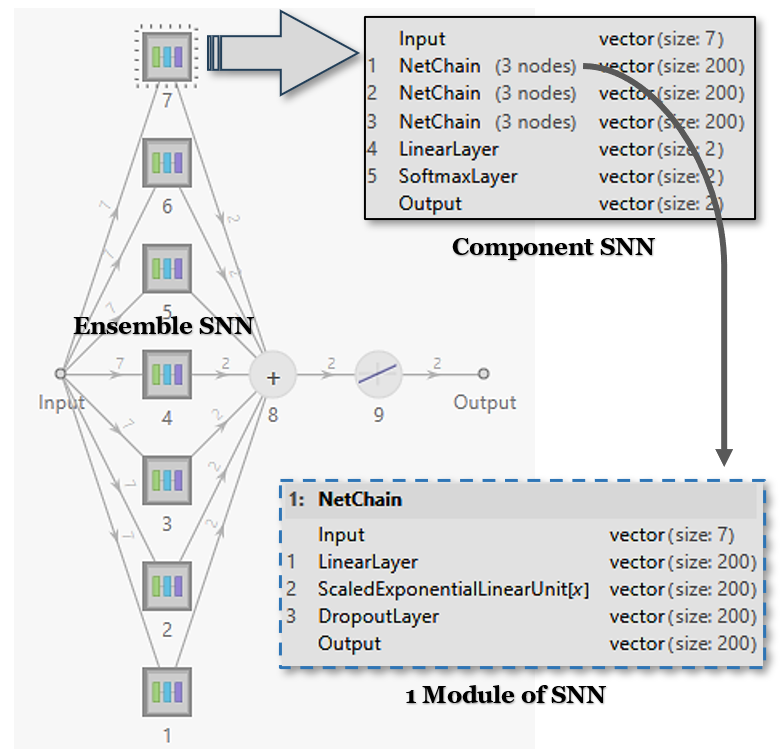}\quad
    \includegraphics[width=0.35\linewidth]{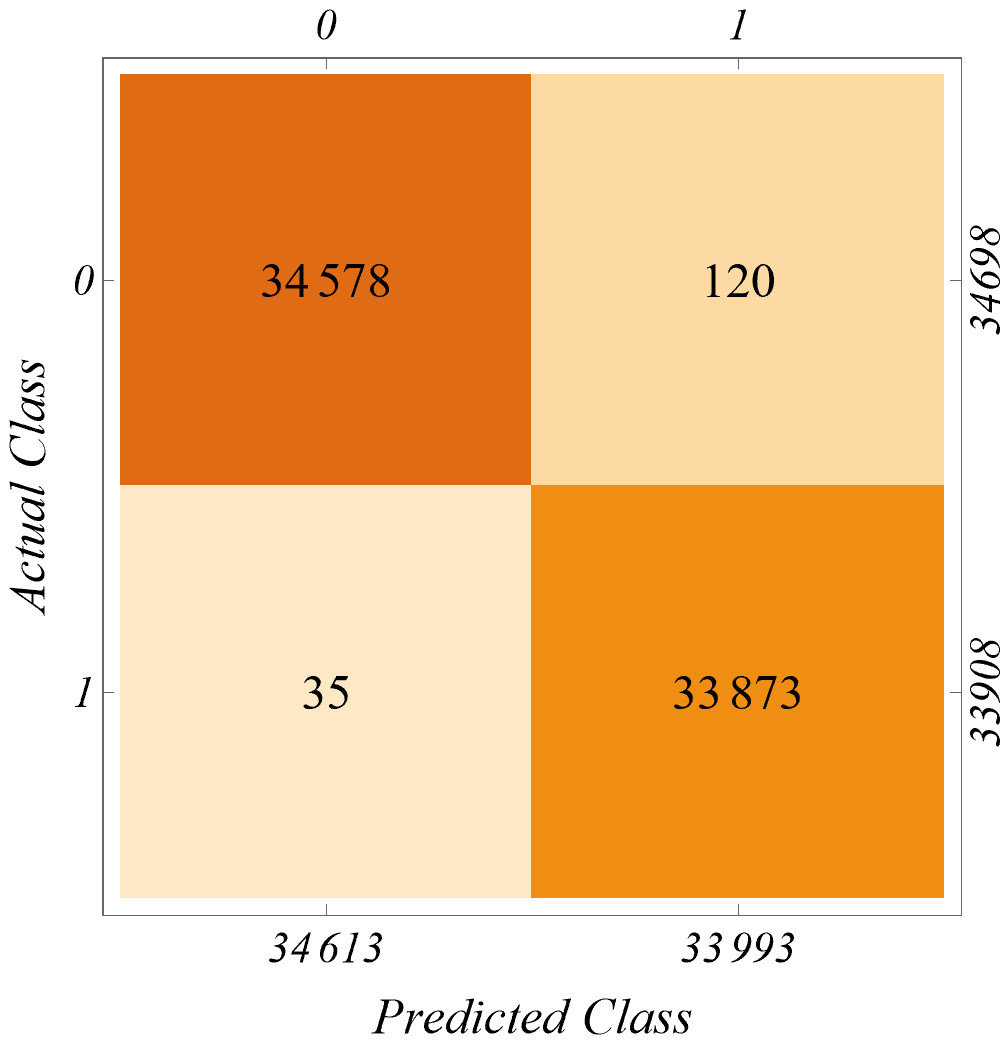}
    \caption{\small Schematic diagram of the ensemble SNN used as a classifier in this work (left) and the confusion matrix (right).}
    \label{fig:SNN}
\end{figure*}

To solve the first problem, we should both run and train the ML algorithms iteratively. The second problem can be solved by using ML predictions to calculate the real likelihoods by running HEP programs, albeit for a much smaller number of points. The solution to the third problem is generating the parameters and theoretical values of observables (or corresponding likelihoods) simultaneously. This can be achieved by training a \textit{generative model} to draw samples from the joint probability distribution $P(\hat{\theta},\hat{O})$ of the parameters $\hat{\theta}$ and the target observables $\hat{O}$. It can then `generate' random instances (outcomes) of these together \cite{mitchell1997machine,10.5555/2980539.2980648}. Any competent generative algorithm can be trained on a vector dataset containing parameters and observables (or $\chi^2$s or $\mathcal{L}$s). 

We thus choose to use some generative algorithm iteratively, using actual evaluation of the likelihoods for an NS as our guide to the final evaluation of the posterior (see figure \ref{fig:flow1}).

\subsubsection{Classification}\label{sec:setup2b} 

After collecting an adequate number of points from the target parameter space that pass the theoretical constraints deriving from vacuum stability and perturbative unitarity \cite{Primulando:2019evb}, we label them either `$1$' or `$0$', depending on whether the corresponding spectrum file is generated or not. To set up an optimal binary classifier, we first use a variable-structure `Self-normalizing Neural Network (SNN)' as shown on the left of Fig.~\ref{fig:SNN}, in a way very similar to that explained in ref.s \cite{Baruah:2024gwy, Bhattacharya:2020vme}. As SNNs can only be used with standardized inputs, a feature extractor function is coupled to the network as an encoder. The hyper-parameters varied are: the number of modules ($n_m$), the number of neurons in each module ($n_w$), and the $\alpha$-dropout probability ($p_{\alpha}$) of the specialized self-normalizing rectilinear unit (SELU). The hyper-parameters were varied over these ranges (initial value, final value, step size): 

\begin{tabular}{l}
     $n_m$ : (4, 10, 2),  \\
     $n_w$ : (50, 200, 50), \\
     $p_{\alpha}$ : (0.007, 0.013, 0.003). 
\end{tabular}

The best combination was found to be $(n_m, n_w, p_{\alpha}) = (4,200,0.007)$, with an accuracy of $99.38\%$ with an ADAM optimizer \cite{kingma2014adam}, which increases to $99.47\%$ with a subsequent training of the same network with a stochastic gradient descent (SGD) optimizer.

We then train an ensemble (7) of nets of that configuration with randomly varied initialization and their predictions averaged over. This ensemble-net has an accuracy of $(99.774 \pm 0.018)\%$ and the corresponding area under the receiver operating characteristic curve (AUC-ROC) is $>99.99\%$. The confusion matrix on the right of Fig. \ref{fig:SNN} shows the false negative rate (FNR or type-II error), i.e, the rate of misidentifying the allowed points as invalid ones, is much lower ($\sim 0.1\%$) than the type-I error ($\sim 0.3\%$), as is need of the hour. This is the classifier we use throughout our analysis. We have trained the final classifier with an incrementally increasing number of data until the obtained accuracy saturates, ensuring that the classifier remains dependable throughout the rest of the analysis.

\subsubsection{Real NVP as a Generative ML model}\label{sec:setup2c} 

Several different Generative Algorithms are at our disposal at present, including but not limited to: Generative Adversarial Networks (GANs) \cite{goodfellow2014generative}, Variational Auto-encoders (VAEs) \cite{kingma2013auto}, and Flow-based generative models \cite{papamakarios2021normalizing,kobyzev2020normalizing}, also called Normalizing Flows (NFs). Neither GANs nor VAEs can learn the probability density of the training data and as will be apparent in this section, it is an integral part of our ML speed-up, although we will not use the learned PDFs directly as our result. We have used a special type of NF, called Real NVP (RNVP; \textit{Real-valued Non-Volume Preserving} transformations) \cite{dinh2016density} in our work. 

Given a dataset of samples collected respectively from a simple enough \textit{base} or \textit{latent} distribution $p_Z$ (easy to collect samples from), and a complex \textit{target} distribution $\hat{p}_X$ of the same degrees of freedom, a normalizing flow learns an invertible and stable mapping between them, represented by an artificial neural network (ANN). As a series of invertible transforms composed together can make a more complex invertible one, carefully designed layers representing the same simple invertible function (with trainable parameters) can thus be used to create the ANN. To find the \textit{normalized} probability density of the target distribution, we then only require the determinant of the \textit{Jacobian} for changing the variables. It is computationally very expensive to calculate the determinant of higher dimensional Jacobians; the requirement of bijectivity (invertibility) makes it even harder. 

RNVP exploits the fact that the determinant of a triangular matrix can be efficiently computed as the product of its diagonal terms. A deep neural network is carefully designed, by stacking constituent layers representing a scalable bijective function (called \textit{affine coupling layers}) together. 
\begin{quote}
    ``In each simple bijection, part of the input vector is updated using a function which is simple to invert, but which depends on the remainder of the input vector in a complex way." \cite{dinh2016density}
\end{quote}
This network can then be used either to calculate the density of the complex target distribution by calculating the density of the \textit{simple} latent distribution or to generate samples from the target distribution by first drawing a sample from the latent one. We use it as a generator.

We use a dataset of 13-component vectors to train the generative algorithm. The first seven are parameters, then the SM-Higgs mass as calculated by \texttt{SPheno} $m_H$, the change in the $\rho$ parameter ($\Delta\rho$, in SM $\rho = 1$), and the non-zero oblique parameters ($\Delta T$, $\Delta S$, $\Delta U$), and finally, the $\chi^2$ given out by HS. Ideally, we should have used all the theoretical predictions for the Higgs sector instead of the last component and compared those to the experimental results. Still, for our purposes, the HS output not only works but also lets us keep track of both theoretical and experimental correlations in this sector with ease.

\begin{figure}[t]
    \centering
    \includegraphics[width=0.9\linewidth]{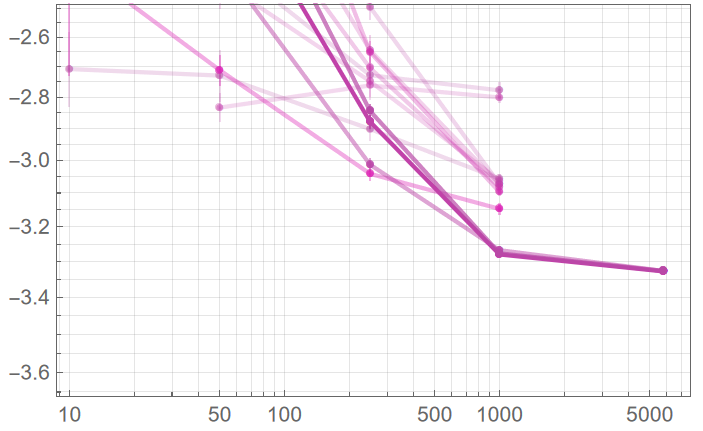}\\
    \includegraphics[width=0.9\linewidth]{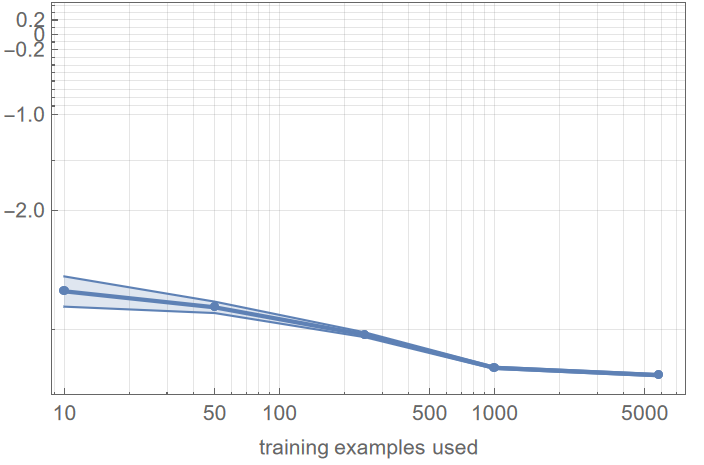}
    \caption{\small Representative learning curve for the training and model selection of the RNVP network. The vertical axis is the Mean Cross-Entropy ($-3.327\pm 0.008$ for the final one for this plot) and the horizontal axis is the log-scaled no. of training examples.}
    \label{fig:learningcurve}
\end{figure}

\subsection{ML-Assisted NS}\label{sec:setup3}
\subsubsection{Initialization and NS}\label{sec:setup3a}

In an NS, the proposal points (some of which sequentially replace the `\textit{live}' points) are sampled from some constrained prior distribution with a specific exploration strategy. The prior in our case is uniform and the exploration strategy is a \textit{Cube}-sampling. The simplest region sampler would use the entire unit hypercube containing the prior, which is prohibitively inefficient. The usual strategy instead is to estimate a gradually shrinking constrained prior as a bounding region of the live points. The multi-dimensional wrapper of this bounding region is generally chosen to be of a specific shape around the \textit{iso-likelihood contour}\footnote{The set of points with the same likelihood, equal to a particular constant. In two dimensions, they constitute a contour line.} defined by the current distribution of live points, i.e., it contains at least the currently estimated volume; in our work that shape is a hypercube. Independent and identically distributed (\textit{iid}) samples are then drawn from that region until one sample passes the current likelihood threshold ($\mathcal{L}>\mathcal{L}^*$).

To keep track of the NS run, we maintain an \textit{object} saving the incremental changes in the \textit{evidence} ($Z$) in addition to the live and `\textit{dead}' (sample) points. At any one point in the run, one may calculate the contribution of the live points to the calculated $Z$ (we will call it the `\textit{Tolerance}' from now on). The run ends only when the tolerance falls below a predetermined threshold ($0.001$ in the present analysis).

\begin{figure*}
    \centering
    \includegraphics[width=0.45\linewidth]{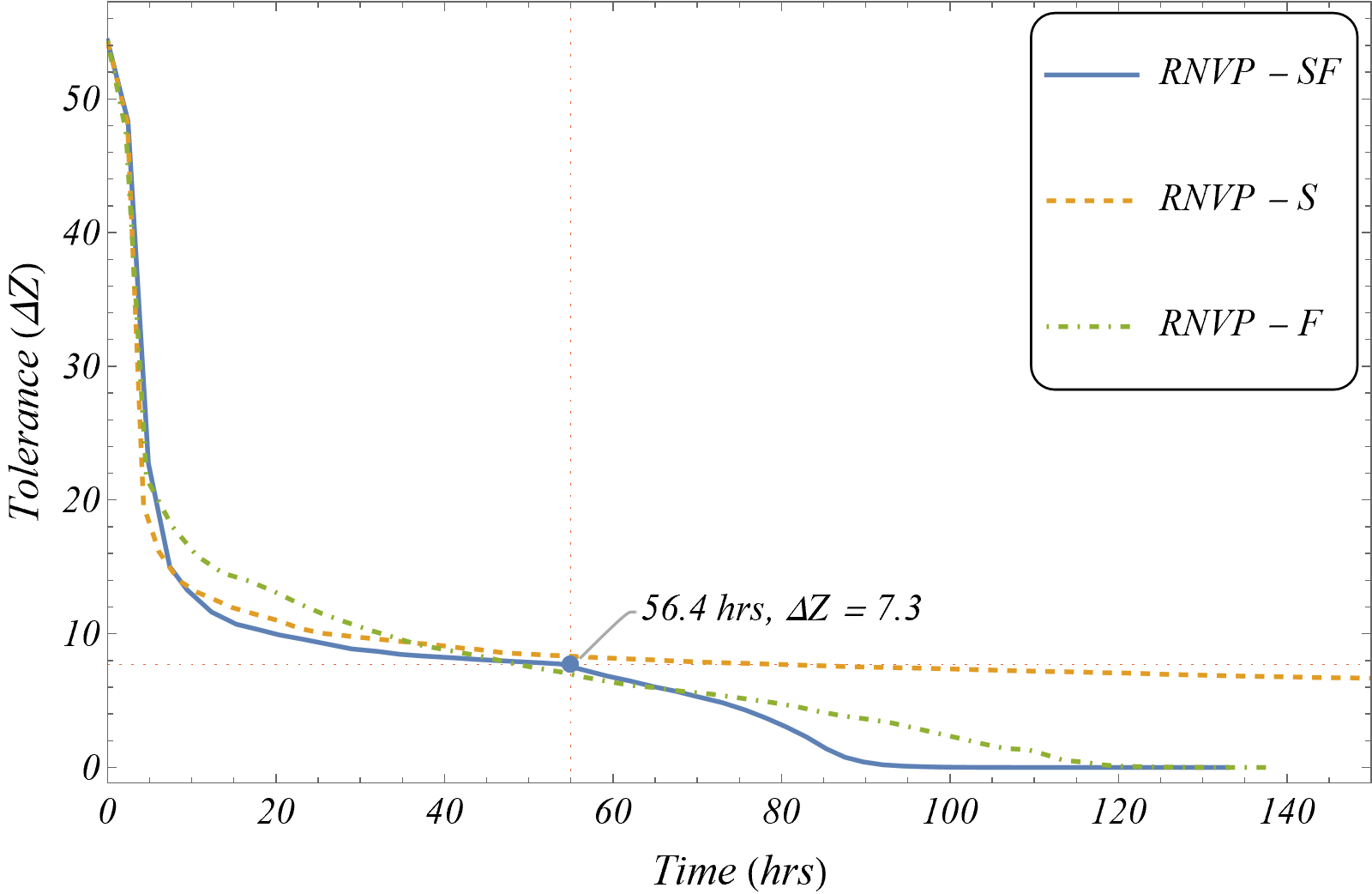}\quad
    \includegraphics[width=0.52\linewidth]{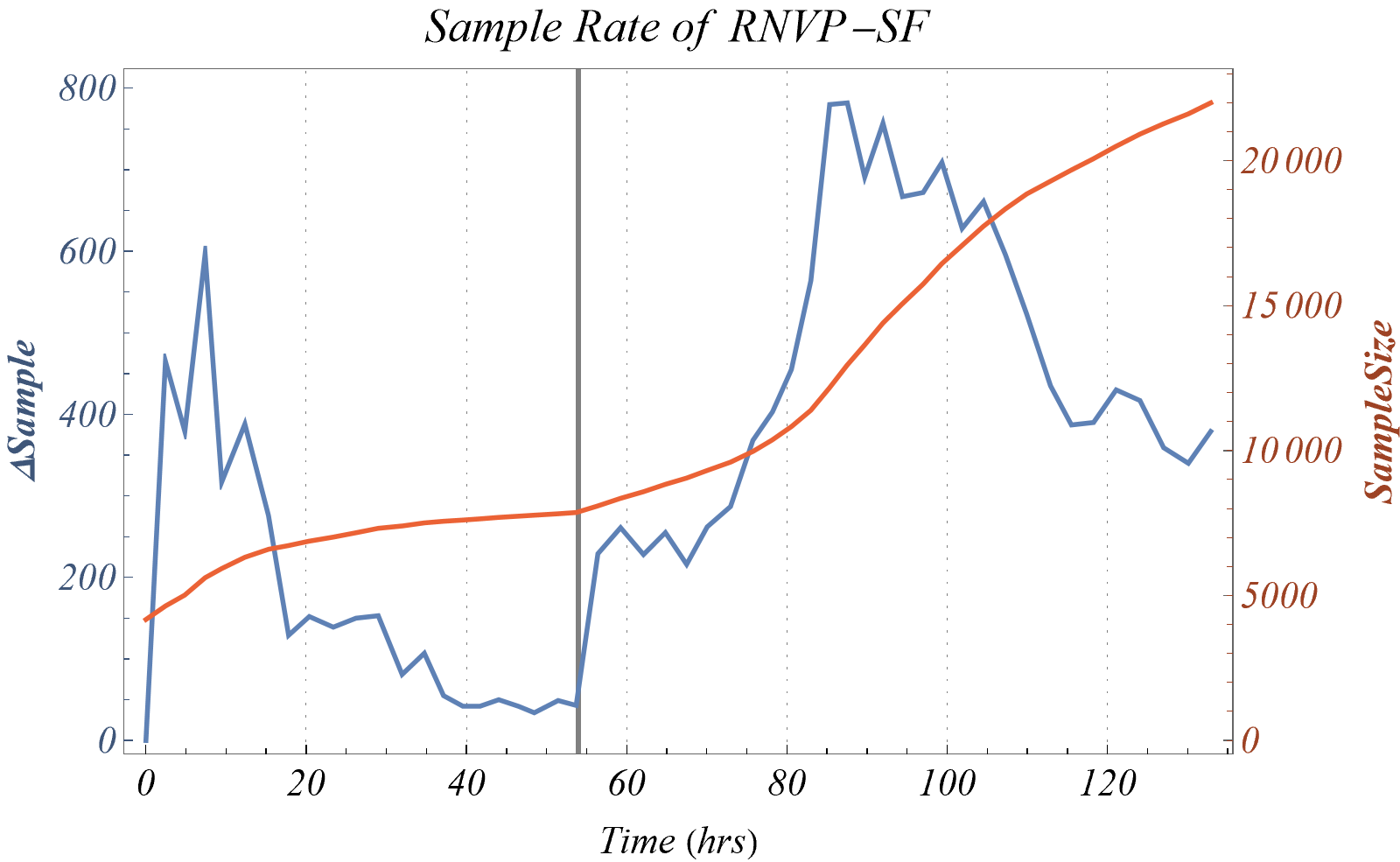}
    \caption{\small Comparison of completion times for different algorithms used in this analysis (left). Time-series plots of (a) NS samples collected at every iteration (blue) and (b) total sample size (red) for $RNVP-SF$ (right). The vertical line depicts the point at which the algorithm switched from \textit{slow} to \textit{fast}.}
    \label{fig:timeCompPlot}
\end{figure*}

After the initial inspection, the starting dataset is prepared with $> 63000$ vectors of 13 components, for which the total $\chi^2$ is not very large ($\leq 1335$). This dataset is used to create a pool of points (labelled `\textit{Collected points}' in figure \ref{fig:flow1}) to train the first iteration of the RNVP generative algorithm, as well as a starting Nested Sample (by using the dataset as a representation of the actual parameter-space). At this starting point, the saved sample only has points sampled from the low-likelihood region and the tolerance is $\approx 50$. 

\subsubsection{RNVP and Training}\label{sec:setup3b}

The \textit{ML-Assisted Nested Sampling} (ML-NS) loop starts at this point. The trained RNVP generator proposes a collection of 13-component vectors sampled from the `Collected points', constrained within the hypercube wrapping the live points. Only some of these remain after applying the theoretical constraints and the classifier on the parameters (the first seven components of any vector). The selected parameters are then passed through the \texttt{SPheno}-\texttt{HiggsTools} combination to generate the 13-component vectors with the actual theoretical observable values (and the actual $\chi^2$ from \texttt{HiggsSignals}). We then use these `\textit{truth}' 13-vectors to check the NS criteria ($\mathcal{L}>\mathcal{L}^*$) and use those that pass the criteria to build the nested sample further. We also add these and a sample of the vectors, failing the criteria, to the `Collected points' pool. Until this point, the classifier does the main speed-up over a simple NS.

The generative performance of the RNVP depends on the distribution of the training samples. A generator, trained only once in the beginning, has a considerably low probability of generating high-likelihood points, as neither the position nor the volume of the high-likelihood region is \textit{a priori} known. The computational expense too, for it to be representative enough, is unknown and probably as large as a naive NS. Even in the rare and ideal case of a large and sufficiently representative starting dataset, using it whole (or an \textit{iid} smaller sample) to train the generator for every iteration would be as prohibitive as using the entire prior for the NS (see sec. \ref{sec:setup3a}).

As the nature of the constrained prior changes non-trivially in exploration problems, the best bet is to iteratively train the generator on a dataset sampled from the constrained prior within the shrinking bounding region. As training is time-consuming, and its predictions do not change considerably for the inner high-likelihood region, it is profitable to train the generator once after several NS iterations. Because we save all the \textit{truth} vectors from the HEP packages in the pooled dataset, the number of samples saved in that pool is a good measure to decide when to train again. We have chosen to train the RNVP network after a group of $6000$ points is dumped in the pool every time. 

To find the RNVP network configuration with the highest possible efficiency, we perform model selection on the two hyper-parameters, namely, \textit{Network Depth} and \textit{No. of Coupling Layers} at every iteration (one representative diagram of the learning curves of this model selection is shown in Fig. \ref{fig:learningcurve}). The final configuration that we find working at almost all iterations is:

\begin{tabular}{ll}
     Activation Function    &:~~ RELU  \\
     Max Training Rounds    &:~~ 750 \\
     Network Depth          &:~~ 2 \\
     No. of Coupling Layers &:~~ 2\,.
\end{tabular}

\begin{figure*}
    \centering
    \includegraphics[width=0.3\textheight]{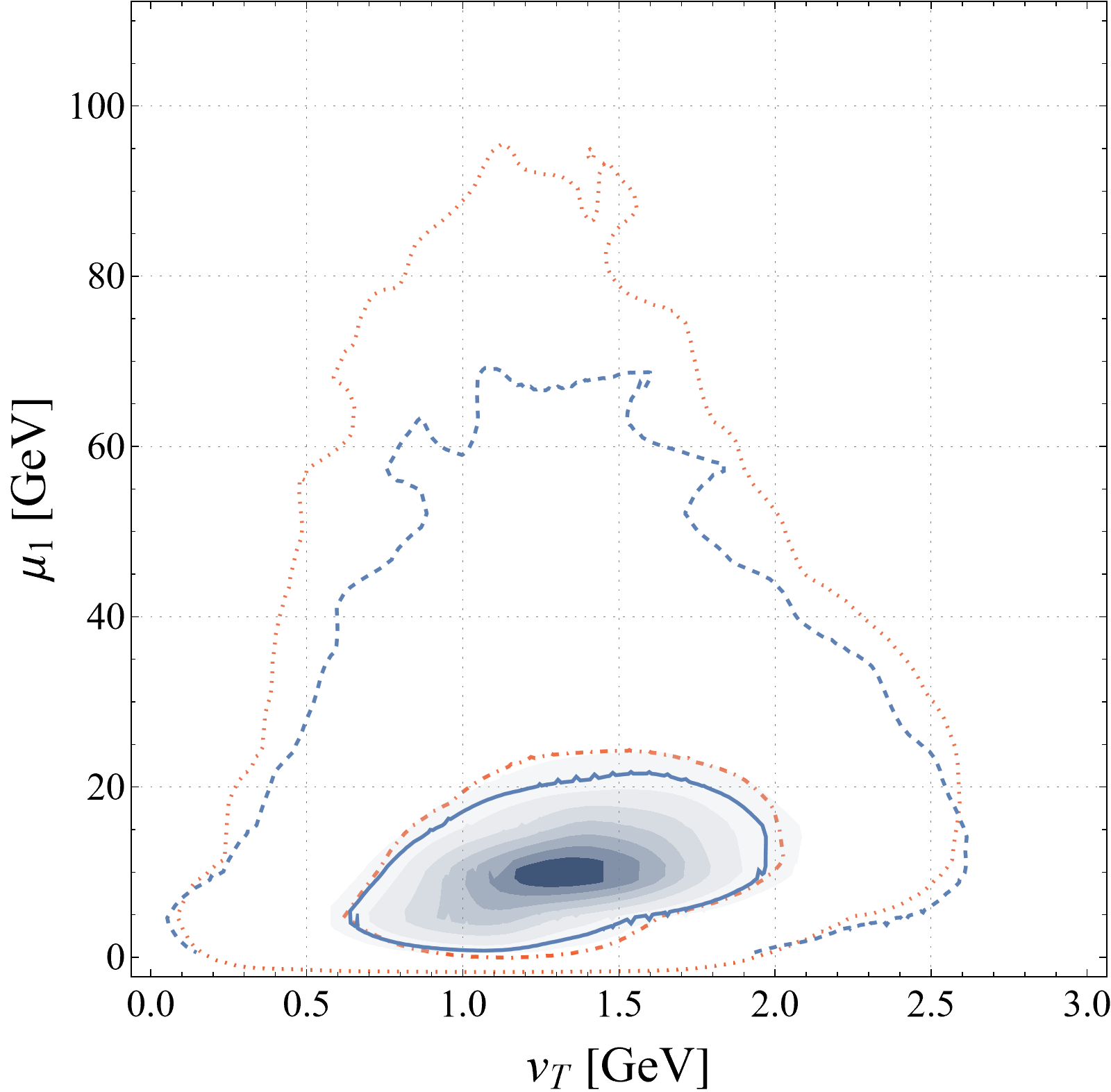}\qquad\qquad
    \includegraphics[width=0.3\textheight]{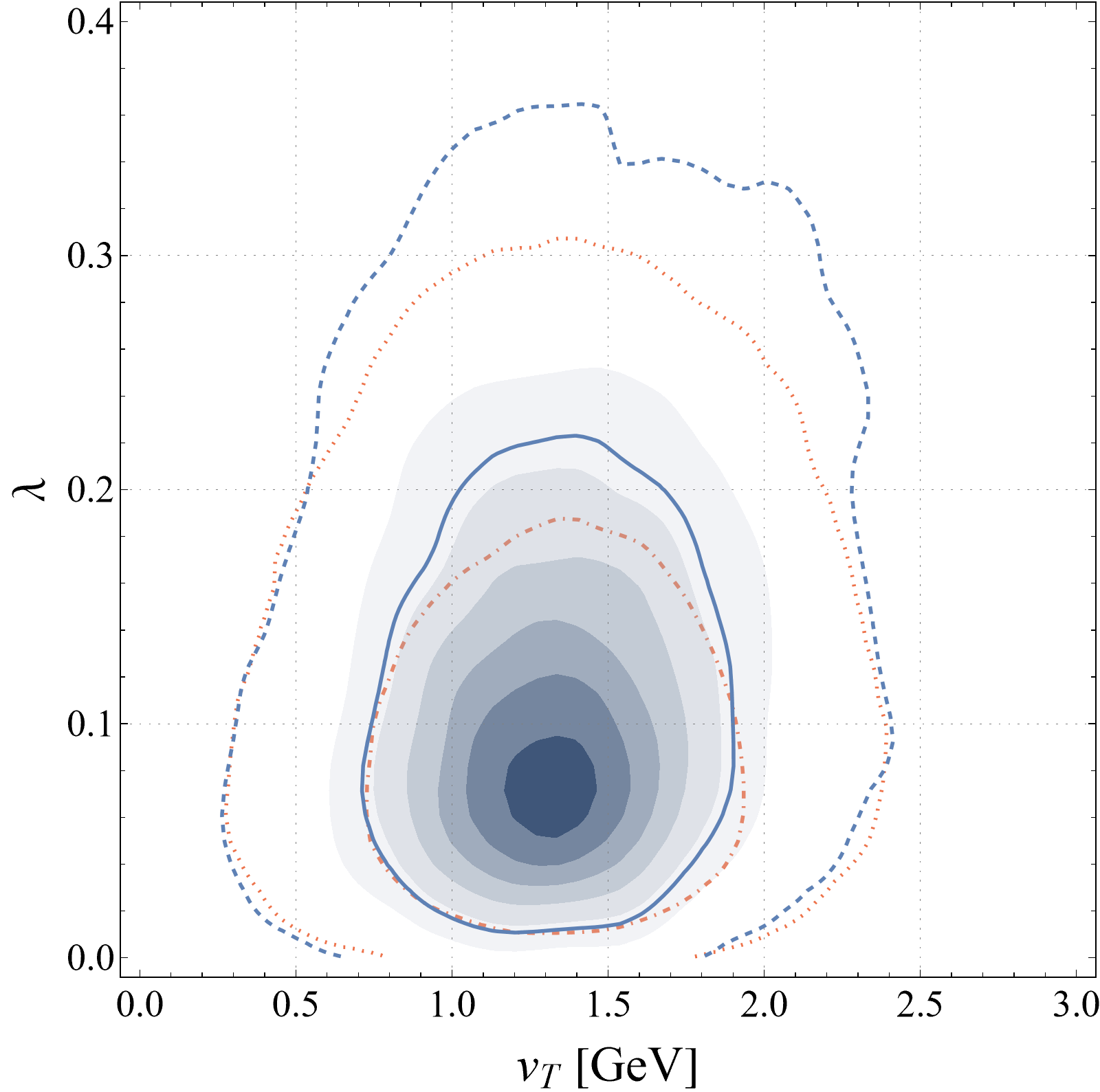}\\
    \vskip 15pt
    \includegraphics[width=0.3\textheight]{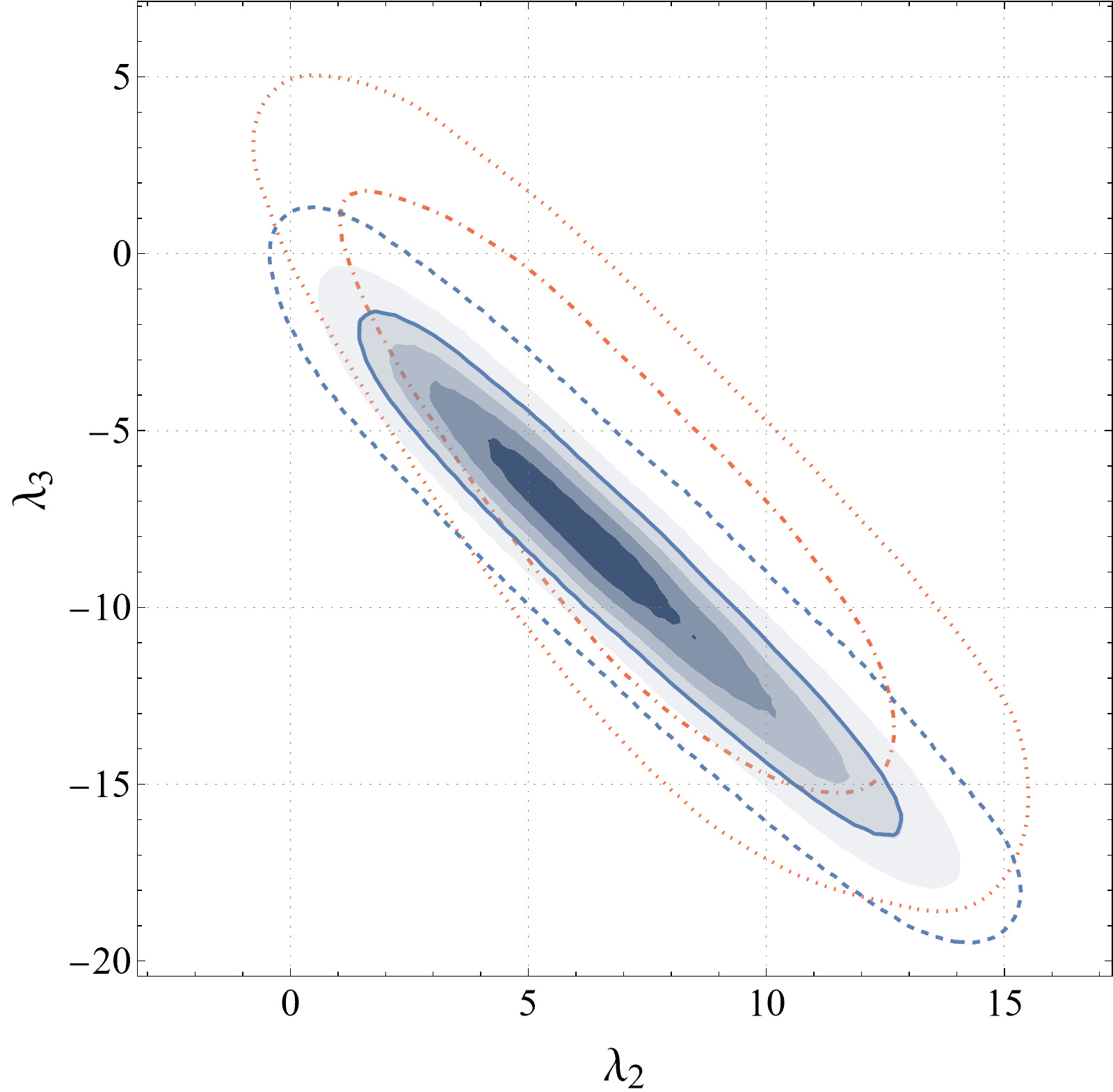}\qquad\qquad
    \includegraphics[width=0.3\textheight]{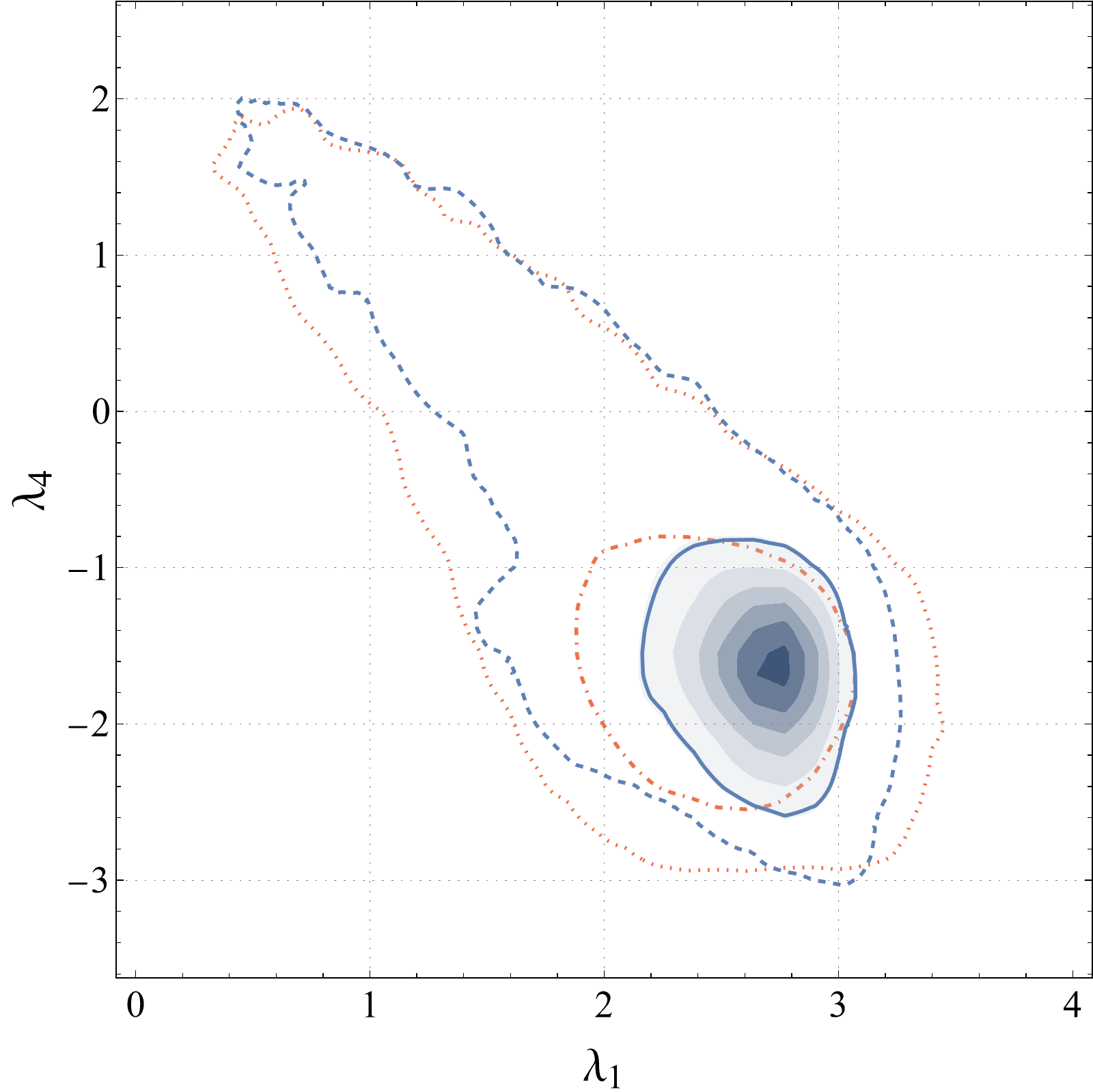}
    \caption{\small Two-dimensional marginal distribution of the posteriors. The contours of the same colour are from the same posterior. The solid and dot-dashed contours correspond to $68\%$ respectively, and the similar $95\%$ credible interval contours are dashed and dotted. Red and blue colours enclose all valid points and those valid points allowed by \texttt{HiggsBounds} respectively.}
    \label{fig:posteriors}
\end{figure*}

\subsubsection{Dataset Selection}\label{sec:setup3b}

The pool (labelled `\textit{Collected points}' in Fig. \ref{fig:flow1}) contains samples from the whole prior. As was discussed in the previous section, predictions from the distribution learned from this become prohibitive. The predictive speed of the generator depends on the way we choose the training dataset from it (labelled `\textit{Database Selection}' in Fig. \ref{fig:flow1}). We have run many instances of the ML-NS algorithm to find the optimal method applicable to any BSM model, of which we exhibit the main three competing options here:
\begin{itemize}
    \item To pick a representative sample from the pool, at every iteration, within the hypercube bounding the \textit{live} points at that instant in the NS-sample, irrespective of their likelihoods. This faithfully represents the prior, but becomes as slow as a naive NS sampler later in the run, when the lowest live-likelihood ($\mathcal{L}^*$) is high, but the constrained prior contains quite a large number of low-likelihood points. Henceforth we call this $RNVP-S$.
    \item To use the iso-likelihood contours with $\mathcal{L}$ slightly lower than the lowest of the \textit{live} points, instead of the bounding region, for dataset selection. When the constrained prior volume is small, it faithfully represents the prior, while being much faster in generating new, high-likelihood points. As we will see, it produces a high bias in the generated sample if used from the beginning, as a result of not correctly representing the prior. We will call it $RNVP-F$ from now on. 
    \item We found in the course of this work that a mixture of the above two methods---$RNVP-S$ at first, then $RNVP-F$ once the tolerance curve (Fig. \ref{fig:timeCompPlot}) flattens enough---not only keeps the sampling unbiased but also does not lose the speed-gain by the ML-assist around the high-likelihood region. Here on we will call it $RNVP-SF$.
\end{itemize}

\begin{figure*}
    \centering
    \includegraphics[width=0.8\textwidth]{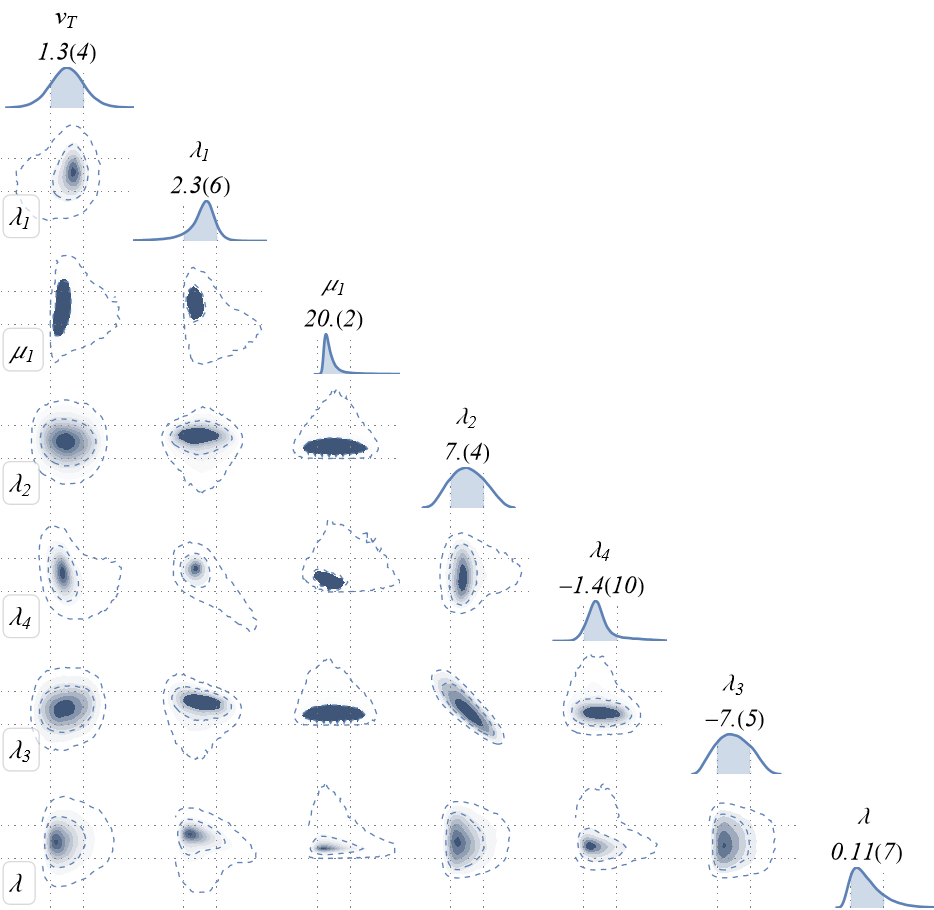}
    \caption{Corner plot depicting all possible marginal distributions and highest likelihood regions of all quartic couplings and triplet VEV.}
    \label{fig:triPlotT2SS}
\end{figure*}

Fig. \ref{fig:timeCompPlot} contains the $\Delta Z$ vs. time curve of three such runs: the orange (dashed) curve represents the $RNVP-S$ one, the green (dot-dashed) curve the $RNVP-F$ and the blue (solid) curve the $RNVP-SF$. The $RNVP-S$ works quite fast at the start but once $\Delta Z$ falls below $10$, it slows down by a lot, adding $3-4$ points to the NS-sample per hour. We truncated the plot here at $150$ hrs., but, in reality, we ran it for $>700$ hrs., with a final tolerance of $\sim 4.09$, when it was finally stopped. While being extremely slow, it was successfully converging to the correct posterior. The $RNVP-F$ run, on the other hand, reached the termination tolerance quite fast and at a steady rate. However, the posterior it generated was wrong (it peaked at a different place than the concentration of the live points, i.e., the high-likelihood region). The first part of the $RNVP-SF$ is identical to $RNVP-S$, but here, when the sample rate (number of NS-samples correctly added between two consecutive trainings) falls below 50 (at $\sim 56$ hrs.) as shown in the plot on the right of Fig.~\ref{fig:timeCompPlot}, we start the $RNVP-F$ procedure. This resulted in a considerable speed-up in the algorithm, without changing the nature of the target posterior perceptibly. The final tolerance reached by both $RNVP-F$ and $RNVP-SF$ is $<0.001$.

\begin{figure*}
    \centering
    \includegraphics[width=0.36\textheight]{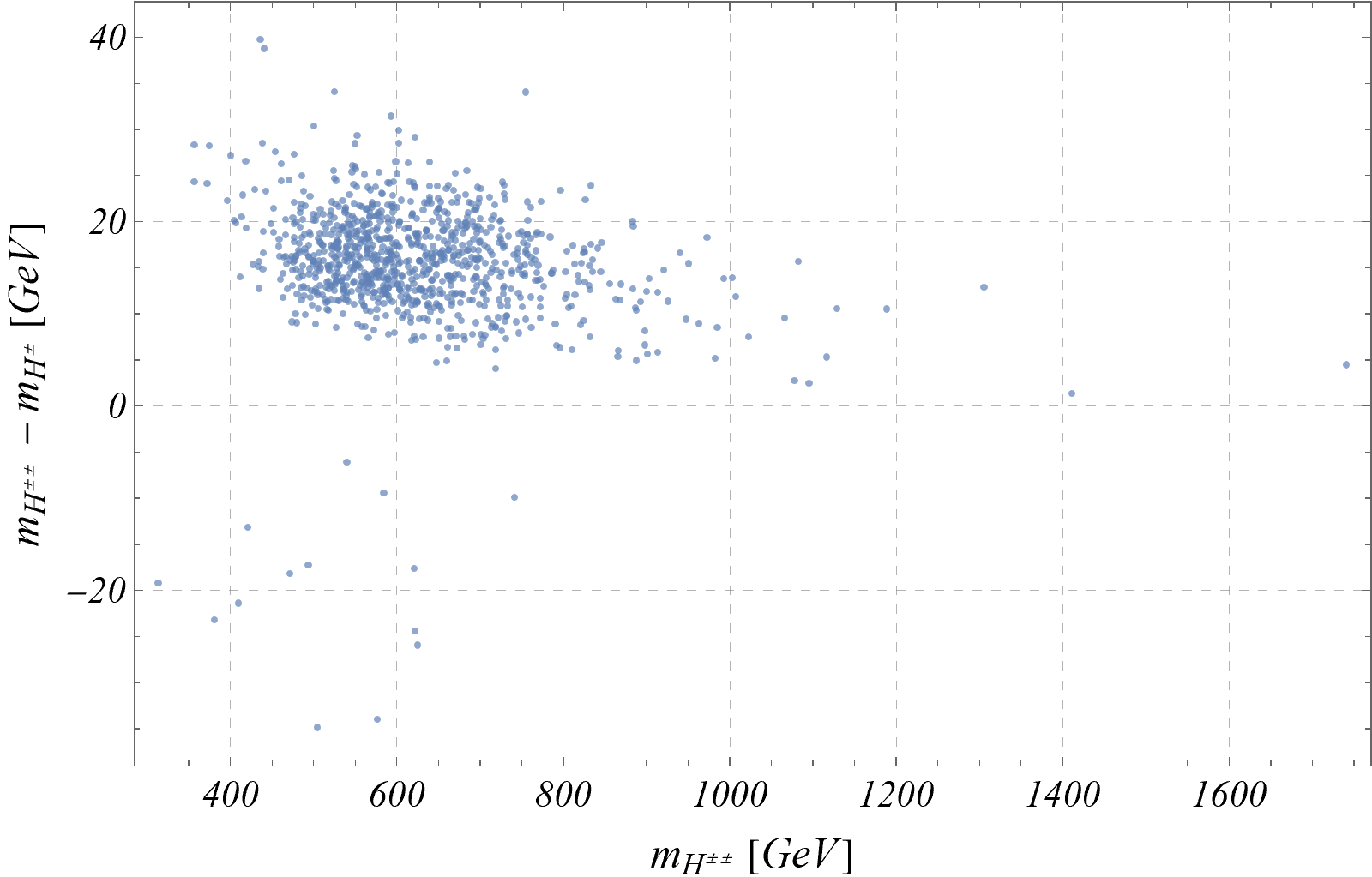}\quad
    \includegraphics[width=0.36\textheight]{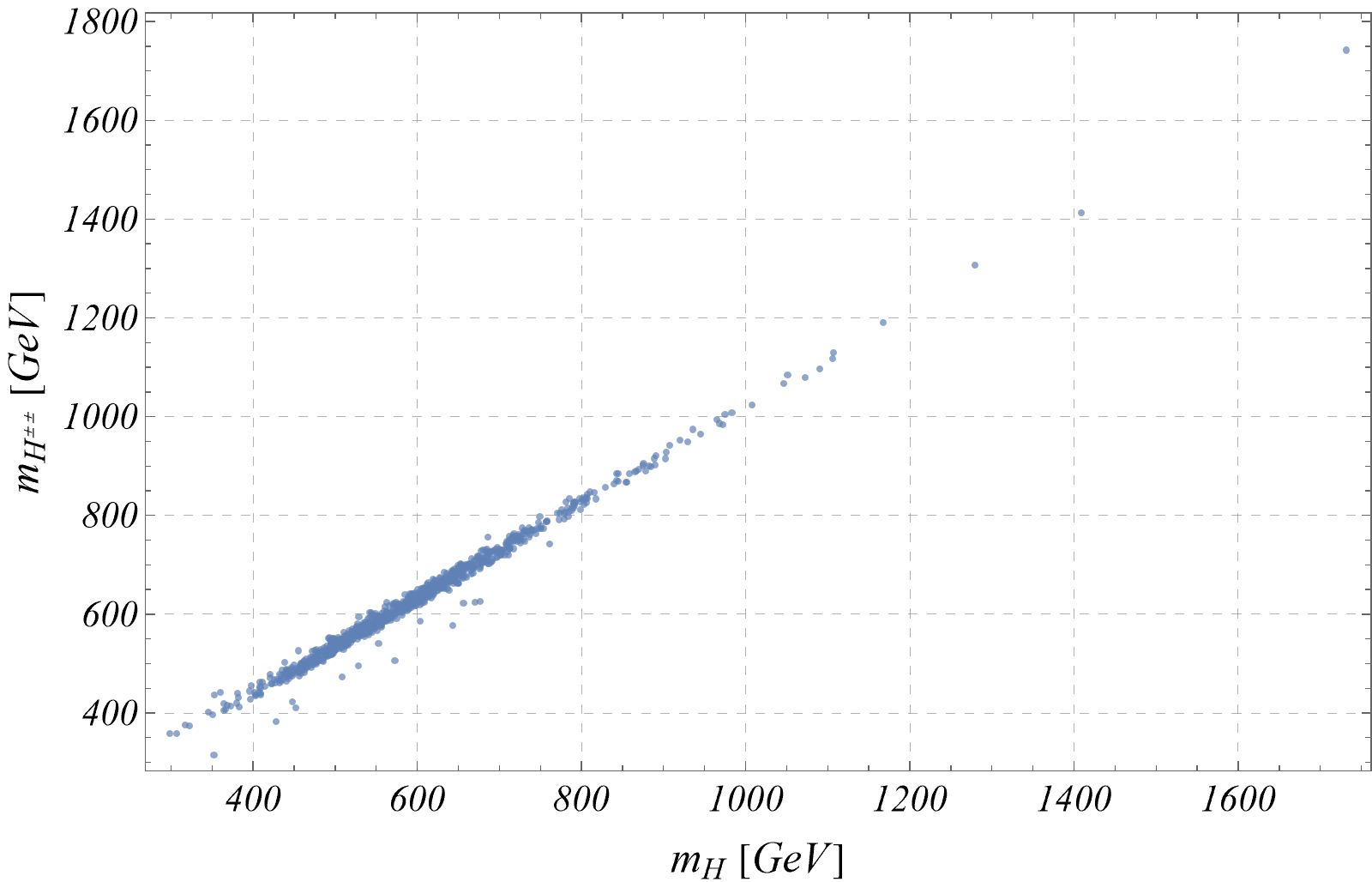}\\
    \includegraphics[width=0.36\textheight]{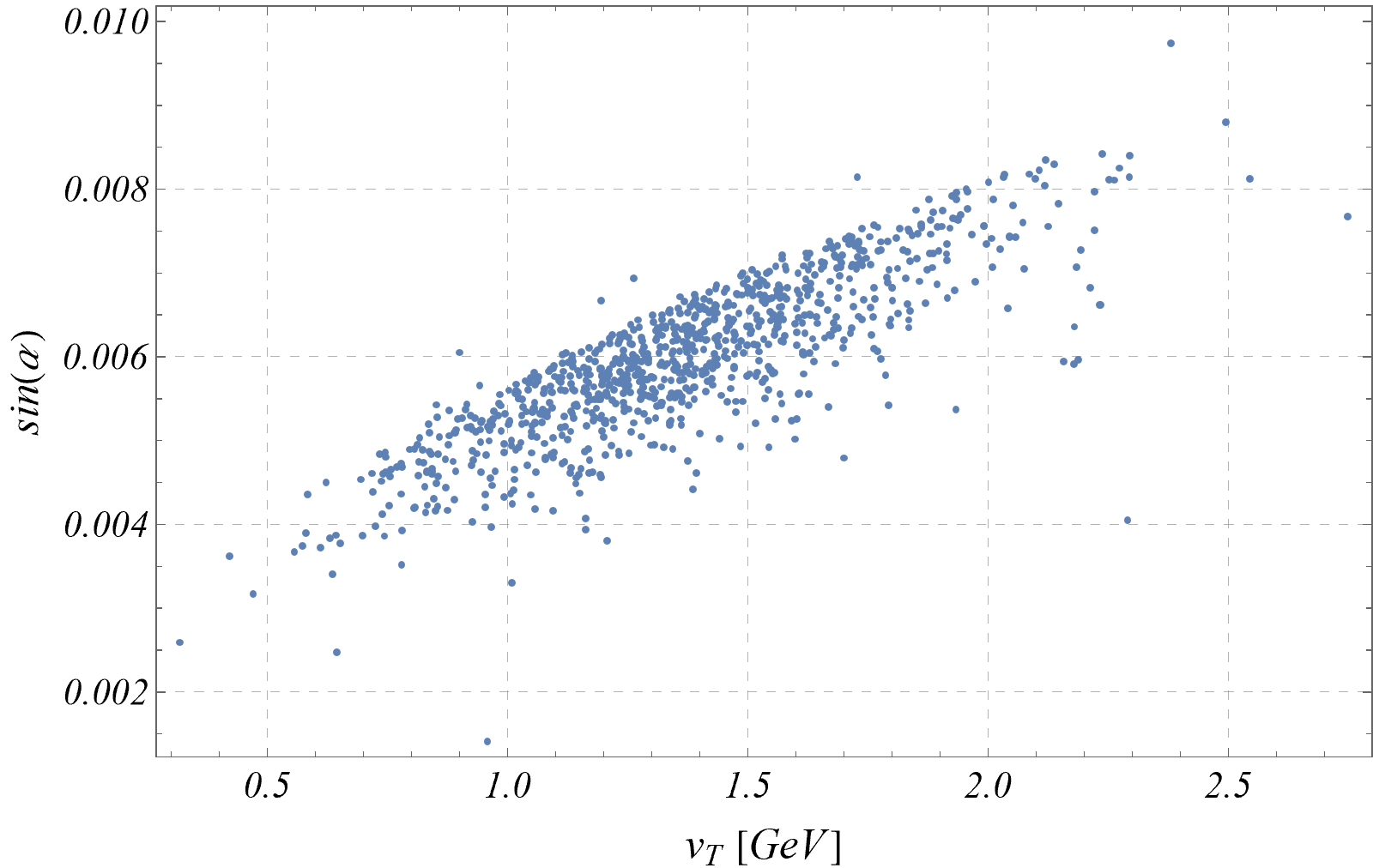}\quad
    \includegraphics[width=0.36\textheight]{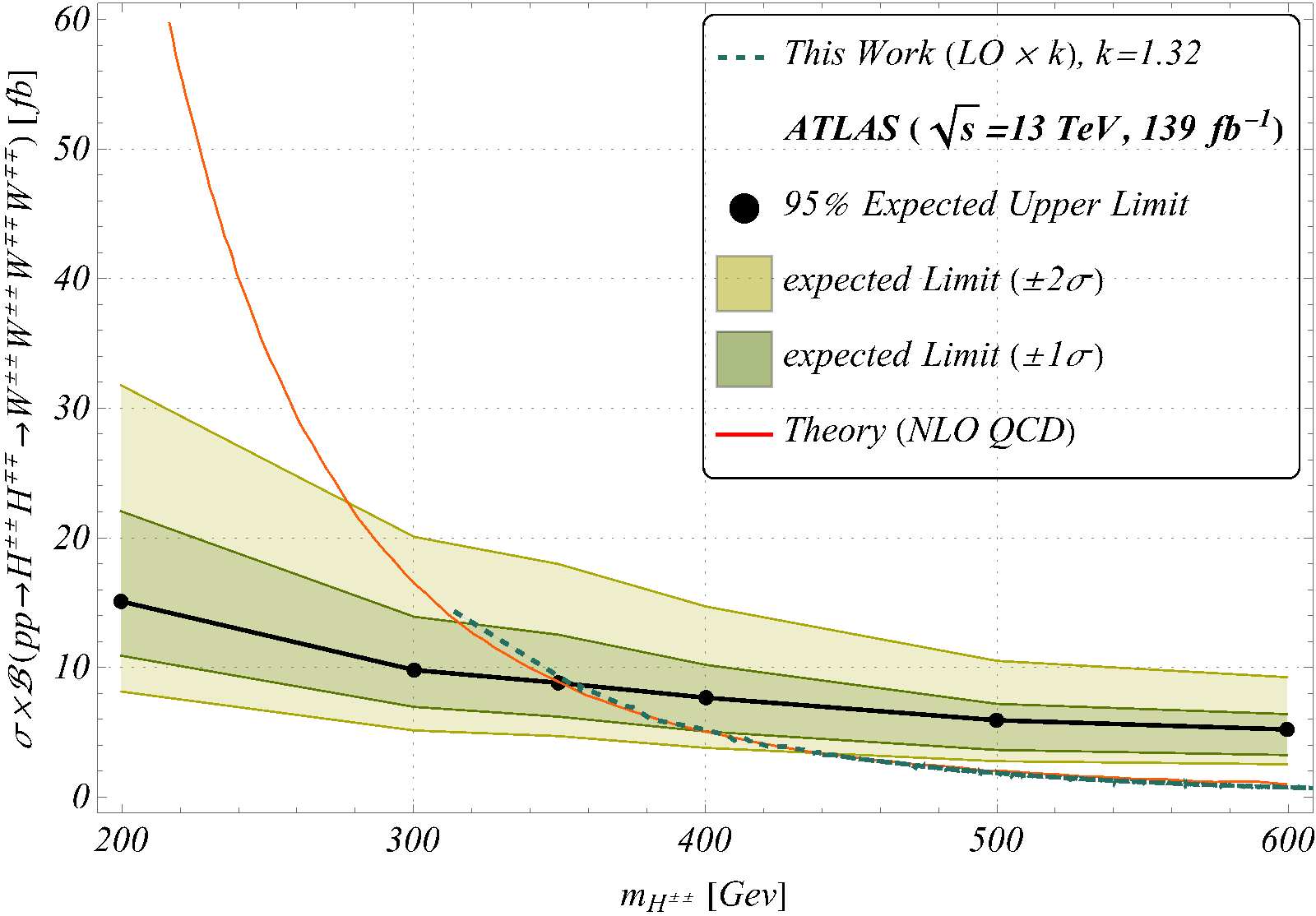}
    \caption{\small All points allowed by all theoretical and experimental constraints up to $95\%$ CL: (a) in the doubly and singly charged Higgs mass plane \textit{(Top-left)}, (b) in the singly charged and neutral heavy Higgs mass plane \textit{(top-right)} (c) The mixing angle between the CP-even Higgs states as a function of the triplet VEV \textit{(Bottom-left)}, and 
    (d) The cross-section $\sigma \times B(H^{\pm\pm} H^{\mp\mp}\to W^{\pm}W^{\pm}W^{\mp}W^{\mp})$ in $fb$ unit as a function of $m_{H^{\pm\pm}}$ along with the $1\sigma$ and $2\sigma$ expected limits \cite{ATLAS:2021jol} \textit{(Bottom-right)}.}
    \label{fig:obsRes}
\end{figure*}

\section{Results}\label{sec:results} 
In this section, we discuss the results obtained through our analysis and highlight their salient features and importance. We have only explored a relatively high $v_T$ region ($\sim 10^{-4} - 2.5$). With this choice, the quartic couplings and the $\mu_1$ parameter were explored to the fullest in this analysis. As mentioned earlier, the objective here was to locate the high-likelihood regions given the present experimental data on the SM-like Higgs boson mass, signal strengths, and the $\rho$-parameter along with the oblique parameters. After locating the favoured parameter space, we have shown the impact of direct scalar (neutral and charged) search results on the obtained parameter space using \texttt{HiggsBounds}. We do the same for the doubly charged scalar with the latest LHC results. All regions of the PS shown in this section obey the vacuum stability and perturbative unitarity conditions. 

\subsection{Posteriors of the parameter space}\label{sec:paraminfo} 
We start with the NP parameters in a mass-agnostic way and determine the posteriors of the PS through our analysis. In Fig. \ref{fig:posteriors} we showcase our results through different two-dimensional ($2D$) marginal distributions. 

The top-left figure shows the distribution in the $\mu_1$ vs. $v_T$ plane. The red contours in the figure indicate the boundaries of the $68\%$ (dot-dashed) and $95\%$ (dotted) credible intervals of the valid sample from the posterior. The blue curves indicate similar boundaries with all points from the sample allowed by \texttt{HiggsBounds} (solid: $68\%$, dashed: $95\%$). The shaded contours with a blue colour gradient paint a more detailed picture of the high-likelihood region (the darker the shade, the higher the likelihood). Thus, the darkest shaded region indicates the most favoured area of the parameter space in this parameter plane. Evidently, any $\mu_1$ values beyond $\sim 70$ GeV is disfavored by the collider data at $95\%$ interval. The triplet VEV is rather tightly constrained with no valid points outside the $0.05 - 2.6$ GeV range subjected to the value of $\mu_1$. The highest likelihood region of $v_T$ however is restricted within $\sim$ 1.2-1.4 GeV. This is mostly due to the $\rho$-parameter. 

The figure on the top right panel of Fig.~\ref{fig:posteriors} shows the $2D$ marginal posterior in $\lambda$ vs. $v_T$ plane. The quartic coupling $\lambda$ is tightly constrained mostly from the SM-like Higgs mass measurement while the triplet VEV on this parameter plane is even more constrained than in the previous figure. This apparent slight change in the credible intervals is a numerical artefact of the $2D$ marginalization of a seven-dimensional space. The \texttt{HiggsBounds} constraints on the obtained sample flatten the marginal modes a little and thus, the recalculated $68\%$ and $95\%$ credible intervals slightly overshoot the corresponding unconstrained (red) contours.

\begin{figure*}[t]
    \centering
    \includegraphics[width=0.36\textheight]{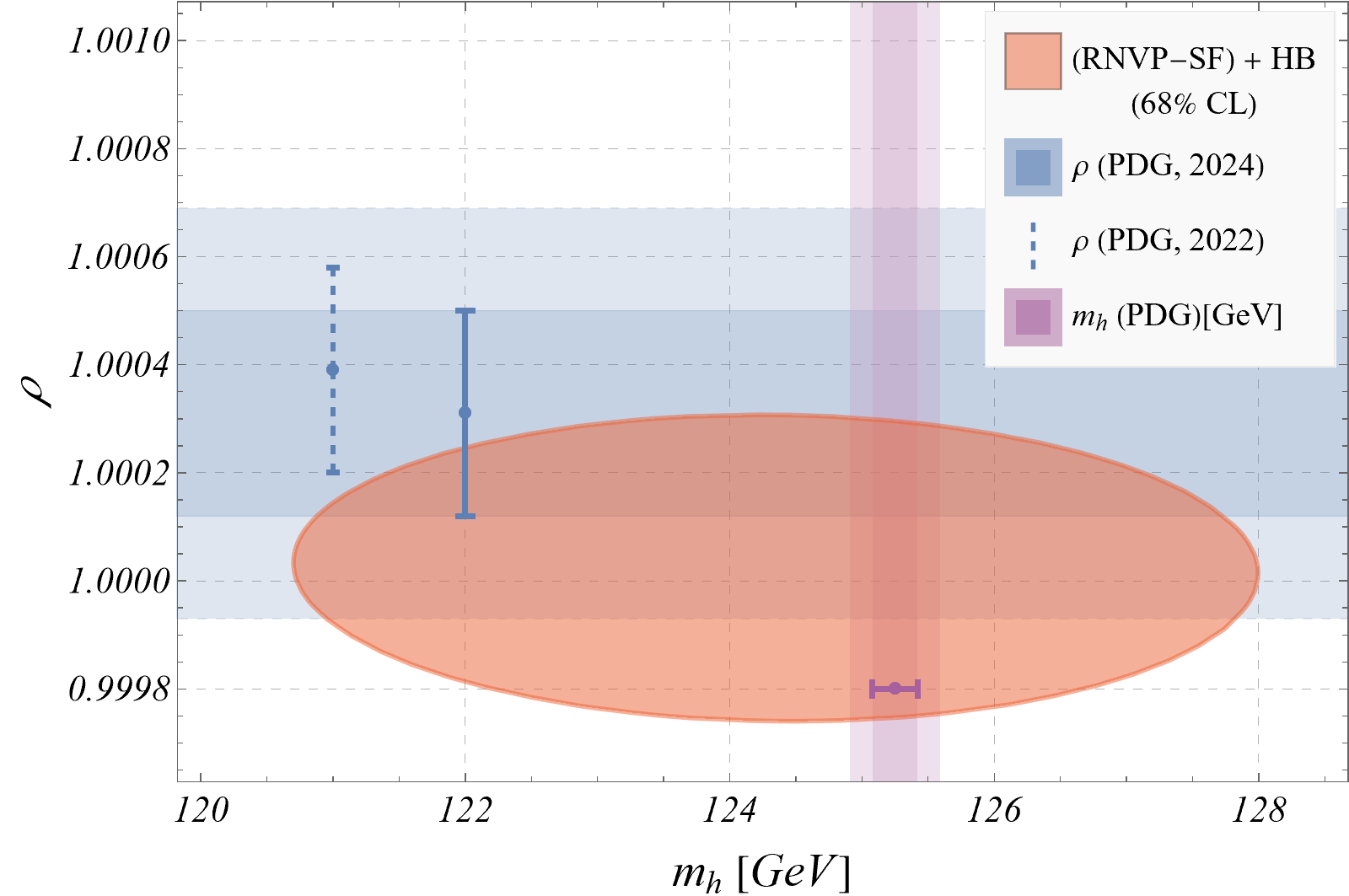}\quad
    \includegraphics[width=0.36\textheight]{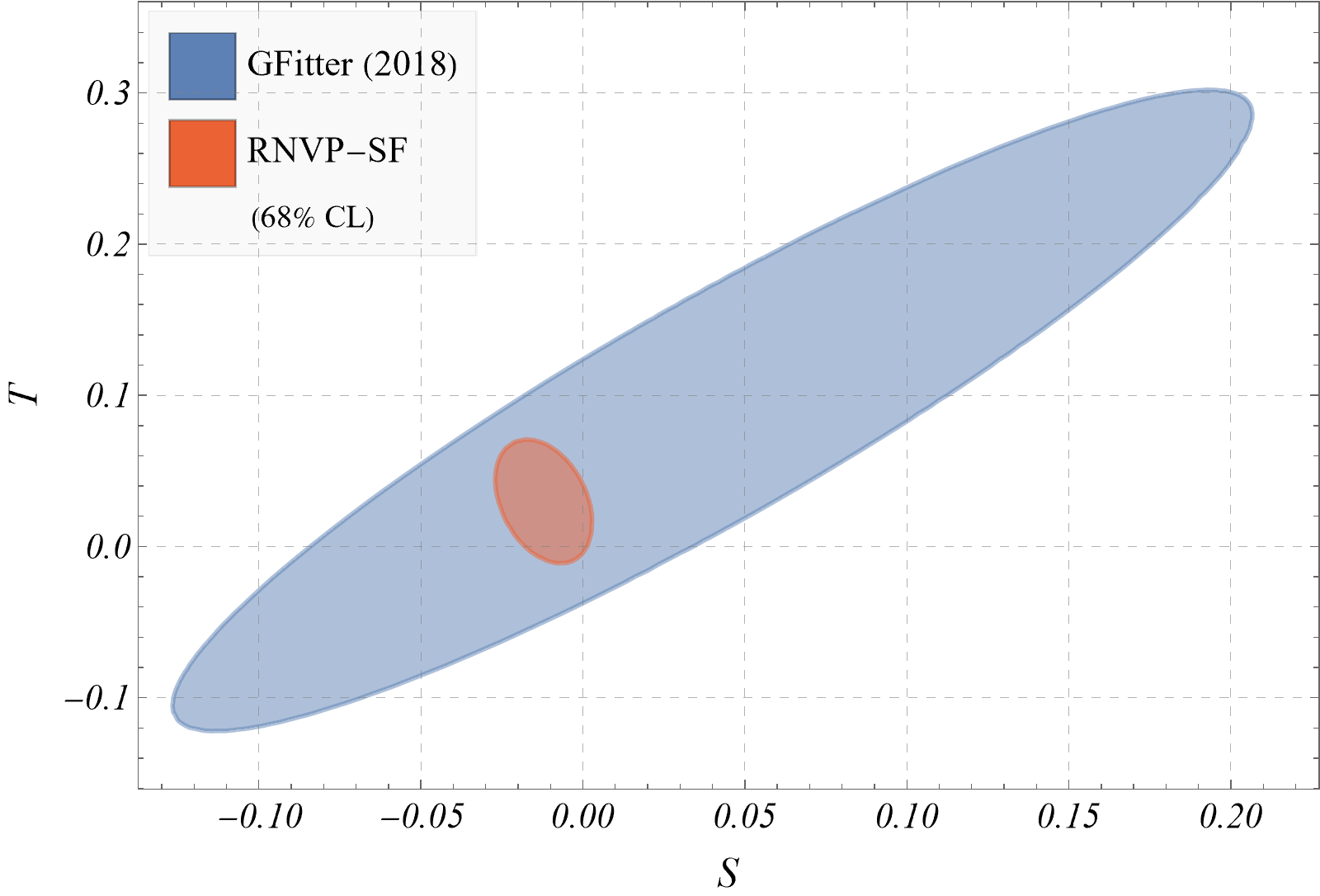}\\
    \caption{\small (Left) Sensitivity of the $\rho$-parameter and the SM-like Higgs mass to the high likelihood region of the parameter space allowed by all constraints. (Right) Comparing the allowed region in the oblique parameters $S$-$T$ plane with that obtained in the high likelihood region from our analysis.}
    \label{fig:mRhoSvT}
\end{figure*}

One can similarly produce and analyze other $2D$ marginal distributions for the rest of the parameters to get a clear picture of PS. The bottom row of Fig.~\ref{fig:posteriors} shows the distributions in $\lambda_3 - \lambda_2$ and $\lambda_4 - \lambda_1$ planes respectively. The plot clearly shows that the correlation between the quartic couplings $\lambda_2$ and $\lambda_3$ is such, that while individually their magnitude can be large, the factor $\lambda_2 + \lambda_3$ always remains small. This mostly follows from the vacuum stability conditions. The bottom-right panel of the same figure shows the $\lambda_4 - \lambda_1$ plane. Apart from the theoretical constraints on $\lambda_1$ and $\lambda_4$, the choices are also restricted from the neutral CP-even scalar mixing angle calculation. For the lighter CP-even Higgs to be SM-like, the mixing angle has to be small and as Eq.~\ref{eq:mix_ang} indicates, the choice of the sum of these quartic couplings depends on the choices of $\mu_1$ and $v_T$. Unlike the other cases, in Fig.~\ref{fig:posteriors} the $95\%$ credible interval contours do not follow the same pattern as their $68\%$ counterparts. The probable reason can be that there exists another high likelihood region with positive $\lambda_4$, which was not detected by our algorithm. One can generate similar posterior distributions with any two input parameters. A corner plot with all possible $1D$ and $2D$ marginal distributions and highest likelihood regions is shown in Fig.~\ref{fig:triPlotT2SS}. The $1D$ plots have their associated $68\%$ credible intervals shown above them. In the $2D$ plots, the dotted contours represent $68\%$ and $95\%$ credible intervals and thus the figure paints a complete picture of the whole parameter space. In Table \ref{tab:paramCorr} we show the correlation among the input parameters obtained from our analysis.

\begin{table}
\caption{The table shows the full correlation matrix derived from our analysis with the seven input parameters.}
\label{tab:paramCorr}
\begin{tabular}
{c|ccccccc}
\hline
   & $v_T$     &  $\lambda_1$ &  $\mu_1$ &  $\lambda_2$ &  $\lambda_3$ &  $\lambda_4$ &  $\lambda$ \\
  \hline
 $v_T$       & 1.    &  0.11    &  $0.05$  &  $-0.01$  &  $-0.18$  &  $0.05$  &  $0.13$  \\
 $\lambda_1$ &       & $1.$     &  $-0.48$ &  $-0.05$  &  $-0.76$  &  $-0.01$ &  $-0.42$  \\
 $\mu_1$     &       &          &$1.$      &  $0.01$   &  $0.27$   &  $-0.02$ &  $-0.03$  \\
 $\lambda_2$ &       &          &          &  $1.$     &  $0.05$   &  $-0.97$ &  $0.02$  \\
 $\lambda_3$ &       &          &          &           &  $1.$     &  $-0.08$ &  $-0.02$  \\
 $\lambda_4$ &       &          &          &           &           &  $1.$    &  $0.08$  \\
 $\lambda$   &       &          &          &           &           &          &  $1.$  \\
   \hline
 \end{tabular}
\end{table}

\subsection{Observable Space}\label{sec:resObs} 
\subsubsection{Physical Masses and properties of the scalars}\label{sec:resScalars} 

With a proper idea of the favoured parameter space at our disposal, let us discuss their impact on the physical scalar masses. As was pointed out in Sec.~\ref{sec:theory}, in addition to one singly- and one doubly-charged scalar, the Type-II seesaw model has one CP-odd and two CP-even neutral scalars. 

The top panel of Fig.~\ref{fig:obsRes} shows the favoured parameter space in different BSM Higgs mass planes. In the top-left figure, the $y$ and $x$ axes show the mass difference between the singly and doubly charged Higgs states and the doubly charged Higgs mass, respectively. Only the points allowed by all theoretical and phenomenological constraints at $95\%$ CL are shown here. It is evident from the figure that most $m_{H^{\pm\pm}}$ values are concentrated in the $\sim 450-800$ GeV range. The mass differences give a clear idea about the corresponding $m_{H^{\pm}}$. Electroweak precision study with the oblique parameters leads to a limit on the mass difference $|m_{H^{\pm\pm}}-m_{H^{\pm}}| \lesssim 40$ GeV \cite{Chun:2012jw}. We obtained the same window for the mass difference without applying this constraint \textit{a priori}, although the points seem to bunch up within the $0-20$ GeV mass gap. The figure on the top-right compares the heavy CP-even Higgs masses ($m_H$) with that of the singly charged Higgs masses. The masses are evidently quite close, lying within a mass gap of $\sim 20$ GeV throughout the favoured PS. This is expected since the mass squared differences $m^2_{H^{\pm}} - m^2_{H}$ and $m^2_{H^{\pm\pm}} - m^2_{H^{\pm}}$ are nearly identical under certain approximations for the range of $v_T$ we have obtained here\cite{Melfo:2011nx}. We have not shown any distribution with the CP-odd Higgs state since it is mass-degenerate with the heavier CP-even state. This is expected since we have not allowed any CP violation in the scalar sector. 
 As is evident from Eq.~\ref{eq:mix_ang}, the mixing angle ($\alpha$) between the two CP-even Higgs states depends on all seven input parameters. The bottom-left panel of Fig.~\ref{fig:obsRes} shows how $\sin\alpha$ varies as a function of $v_T$. Understandably, $\sin\alpha$ has to be small owing to the present-day Higgs data and the figure clearly shows that it can be at most $\sim 10^{-3}$. All points in this figure are allowed by all theoretical and phenomenological constraints at the $95\%$ CL. Note, that any smaller mixing angle, though allowed in principle, will not fit the SM-like Higgs data within $95\%$ CL. 


\begin{figure*}
    \centering
    \includegraphics[width=0.8\textwidth]{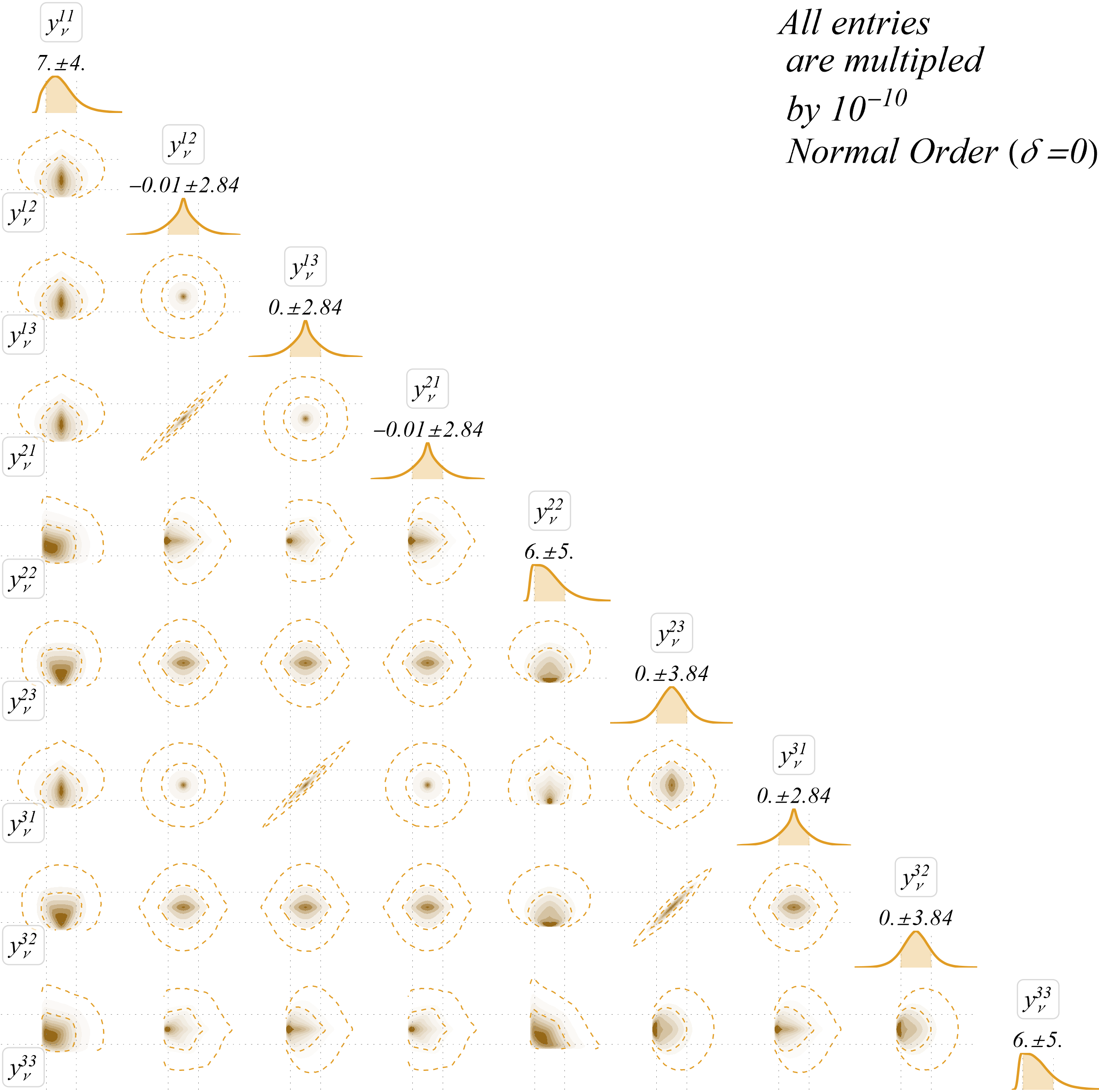}
    \caption{Corner plot depicting all possible marginal distributions and highest likelihood regions of all neutrino Yukawa matrix elements assuming NO in light neutrino masses. The phase angle $\delta = 0$. }
    \label{fig:grYuDeloNO}
\end{figure*}

Note that, all scalars apart from the doubly charged Higgs state have already been subjected to direct search constraints through \texttt{HiggsBounds}. To comment on the impact of direct search results of doubly charged Higgs states on the favoured parameter space obtained here, we have to first talk about possible decays of $H^{\pm\pm}$. Since we are working with a large $v_T$ scenario, the Yukawa couplings have to be extremely tiny to be consistent with neutrino data, as indicated by Eq.\ref{eq:neut_mass}. All LNV decays are, therefore, highly suppressed over all of the favoured parameter space. As a result, the only possible two-body decay mode of this scalar is $H^{\pm\pm}\to W^{\pm}W^{\pm}$ which dominates over all possible three-body decay modes. The other possible two-body decays e.g., into $H^{\pm}H^{\pm}$ or $H^{\pm}W^{\pm}$ are not kinematically allowed due to the small mass gap of $m_{H^{\pm\pm}}$ and $m_{H^{\pm}}$. We have checked the consistency of the favoured parameter space against the latest collider search results of doubly charged Higgs; produced via vector boson fusion (VBF)\cite{ATLAS:2024txt,ATLAS:2023dbw,CMS:2021wlt} or pair-produced via photon or $Z$-boson mediated diagrams \cite{ATLAS:2021jol}. They decay into $W^{\pm}W^{\pm}$ in both cases. The cross-section was computed using \texttt{MadGraph} \cite{Alwall:2014hca,Frederix:2018nkq}. We find that for the VBF production, the whole favoured PS is quite safe from the existing results with the cross-section $\sigma _{\rm VBF}(H^{\pm\pm})\times(H^{\pm\pm}\to W^{\pm}W^{\pm})$ lying at least one order of magnitude below the experimental limit. The doubly charged Higgs can also be pair-produced at the LHC with both of them decaying into $W$-bosons \cite{ATLAS:2021jol}. The experimental collaboration studied the experimental observation in the context of the Type-II seesaw model and showed that $m_{H^{\pm\pm}}\lesssim 350$ GeV is excluded from this search at 13 TeV centre-of-mass energy with the luminosity at $139 {\rm fb}^{-1}$ luminosity. Almost all the points allowed by theoretical and phenomenological considerations up to $95\%$ CLs produce $m_{H^{\pm\pm}} \gtrsim 400$ GeV with a sparse region in between $350-400$ GeV. We show the comparison with the experimental result on the bottom-right of Fig.~\ref{fig:obsRes}. We have shown the $1\sigma$ and $2\sigma$ expected limit and the $95\%$ expected upper limit for $\sigma \times BR(pp\to H^{\pm\pm}H^{\mp\mp}\to W^{\pm}W^{\pm}W^{\mp}W^{\mp})$ along with the theoretical prediction obtained for Type-II seesaw model \cite{ATLAS:2021jol}. The red line represents the cross-section calculated at next-to-leading-order by the experimental collaboration and the dotted black line shows our leading-order calculation multiplied by a k-factor of 1.32. We can safely conclude from this figure that the favoured parameter space is yet to be explored by the collider search of $H^{\pm\pm}$.

\subsubsection{Yukawa matrix from neutrino oscillation data}
We have only discussed the scalar sector up to this point. Once the triplet VEV in known, one can determine the neutrino Yukawa matrix using Eq.~\ref{eq:neut_mass} such that the light neutrino masses and mixing angles are consistent with the neutrino oscillation data \cite{deSalas:2020pgw}. We have performed the calculation for both normal (NO) and inverted ordering (IO) of light neutrino masses. The neutrino oscillation data is made up of two mass-square differences, three mixing angles, and a CP-violating phase angle ($\delta$). 

Eq.~\ref{eq:YukMat} shows the resultant neutrino Yukwa matrix elements, subjected to neutrino oscillation data and for the choice of $v_T\sim 0.5 - 2.5$ GeV favoured by the constraints at $95\%$ CL from our analysis. 
\begin{eqnarray}
Y=\begin{pmatrix}7.\pm 4. & -0.01\pm 2.84 & 0.\pm 2.84 \\
-0.01\pm 2.84 & 6.\pm 5. & 0.\pm 3.84 \\
0.\pm 2.84 & 0.\pm 3.84 & 6.\pm 5.
\end{pmatrix}\times 10^{-10}  \nonumber \\
\label{eq:YukMat}
\end{eqnarray}
We have kept $\delta=0$ in this calculation and have assumed NO in neutrino masses. The hierarchy is clearly reflected in the resultant Yukawa elements. The central tendencies (mean) and associated dispersions (standard deviation, $\sigma$) corresponding to all elements are quoted. As expected, given the favoured region of $v_T$, the Yukawa elements are all $\sim 10^{-10}$ with even smaller off-diagonal entries. This ensures that all LNV or LFV decays are well suppressed. Note that, one can perform a similar calculation with a non-zero $\delta/\pi = 194 ^{+24}_{-22}$ \cite{deSalas:2020pgw}. The phase angle will require the Yukawa matrix to be complex. However, the large uncertainty associated with it results in large, flat posteriors, with overlapping $68\%$ and $95\%$ regions. We show posterior distributions of the Yukawa elements for the NO scenario with $\delta=0$ in Fig.~\ref{fig:grYuDeloNO}. One can generate similar corner plots for the IO scenario and with non-zero $\delta$. These results, along with all other results (plots, samples, and trained networks), are available in the \href{https://github.com/sunandopatra/MLNS-T2SS}{GitHub repository}. 

\begin{figure}[t]
    \centering
    \includegraphics[width=0.9\linewidth]{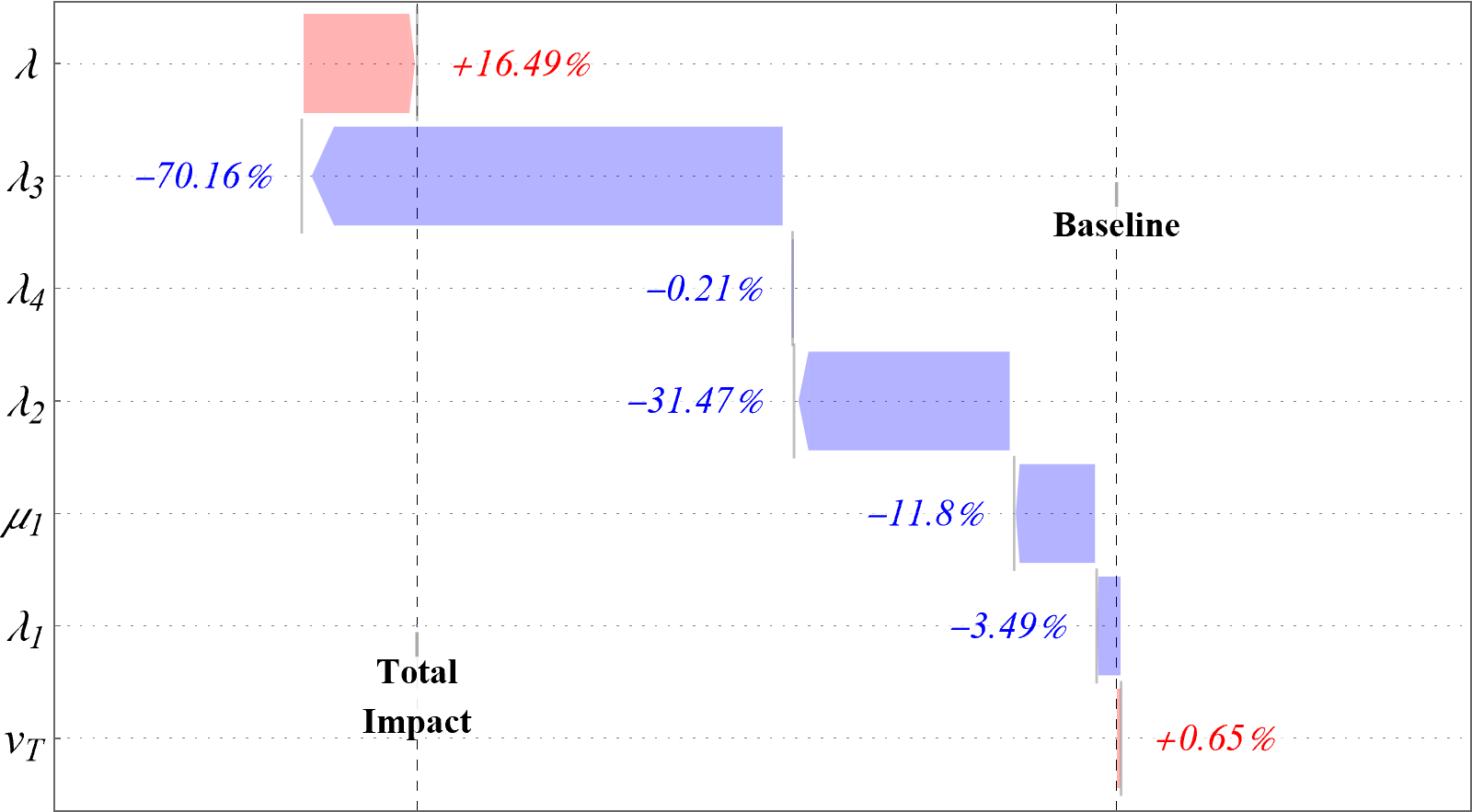}\\
    \vskip10pt
    \includegraphics[width=0.8\linewidth]{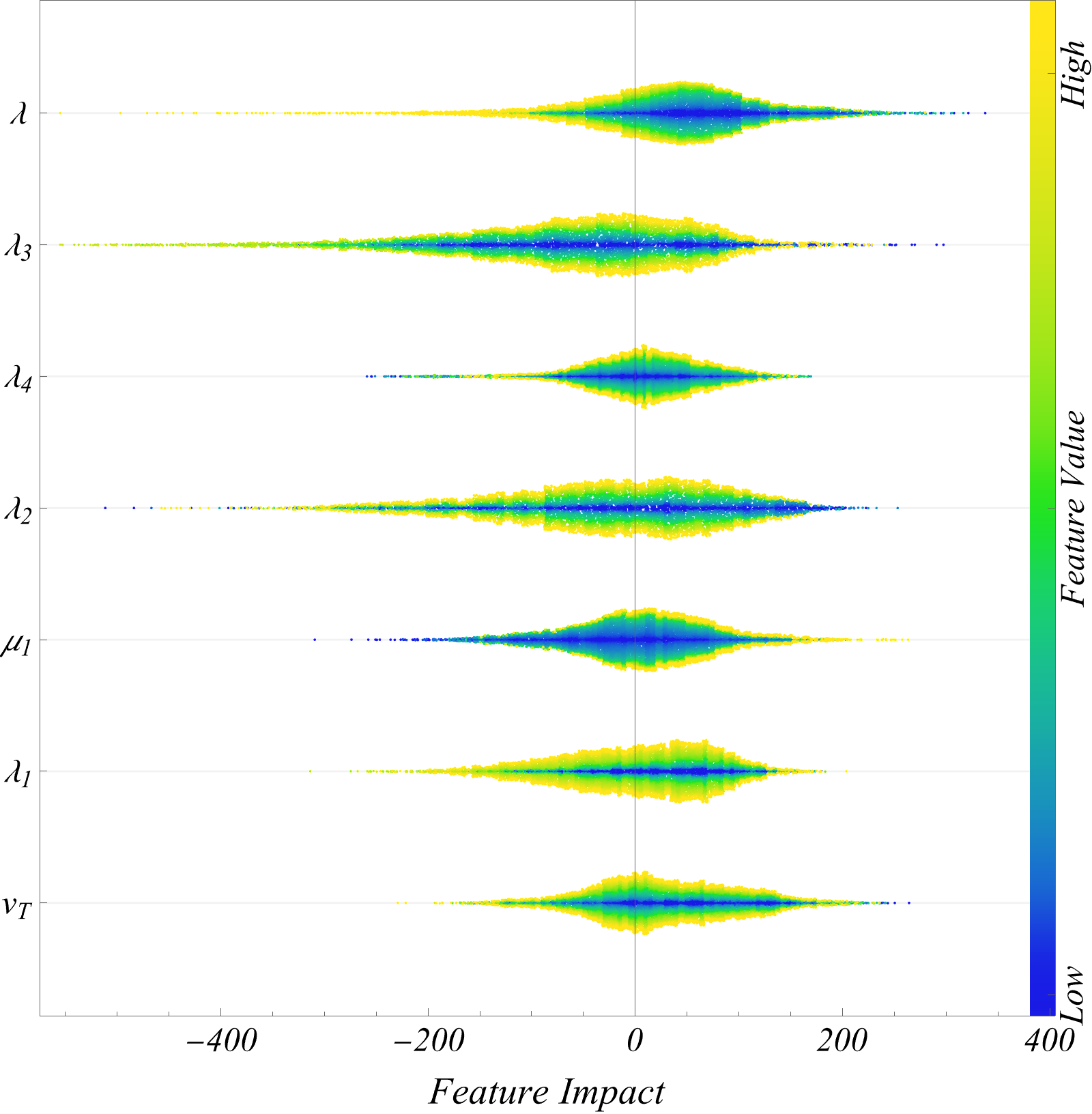}
    \caption{\small \textit{Waterfall Plot} demonstrating the additive nature of the SHAP at a high-likelihood point (above) and the distribution of the Shapley values over the whole dataset (below).}
    \label{fig:shap}
\end{figure}

\subsubsection{Sensitivity of the Observables to the parameters}

In Fig.~\ref{fig:mRhoSvT} we show the sensitivity of some precisely measured observables to the obtained parameter space. To get the distribution of the observables in these plots, the whole posterior---allowed by all theoretical and experimental constraints---is used and a Bi-normal distribution is estimated. On the left, we show the overlap of the observable distribution (the red ellipse showing $68\%$ credible interval) calculated from the posterior with the $68\%$ and $95\%$ experimental ranges of $\rho$ parameter and the SM-like Higgs mass $m_h$. On the right, we show the equivalent region on the $S-T$ plane (red ellipse) and compare it with the $68\%$ CL computed by the GFitter group \cite{Haller:2018nnx} and it is clear that the high likelihood region obtained through our analysis is well within the experimental electroweak precision bounds. Regarding the $\rho$-parameter, we started our analysis with the estimate made by the Particle Data Group (PDG) in 2022 which was $\rho=1.00039\pm 0.00019$ \cite{ParticleDataGroup:2022pth}. While we were performing our analysis, PDG updated their estimate which now is quoted as $\rho=1.00031\pm 0.00019$ \cite{ParticleDataGroup:2024cfk}. It does not affect our analysis or results. In fact, as the figure shows, the new measurement is in even better agreement with our estimated parameter space.

To check the sensitivity of our analysis to the parameters, we take the help of the `\textit{SH}apley \textit{A}dditive ex\textit{P}lanations' (SHAP) \cite{SHAP1}. It is a widely used method to calculate the relative importance of features (parameters, for our case) in obtaining a prediction (observables/likelihood for us), using game-theoretic approaches. To use this in our analysis, we first use our final sample from posterior (allowed by \texttt{HiggsBounds}, $>93000$ points) to obtain the respective negative log-likelihoods, which together make both training and test sets. After successfully training a \textit{Gradient Boosted Trees} algorithm (we could have used some sophisticated network as well, but the SHAP would hardly change), we apply it to the test dataset to obtain the predictions and corresponding measurements. The ML algorithm is just a placeholder here for a non-linear predictor function. We do the SHAP analysis with these.

A vector of Shapley values is calculated for every point prediction, with one component for each of the features. These values signify the relative importance of the respective feature in obtaining that prediction using both the impact of the change in that feature on the output, and the distribution of that feature. With one of the features fixed at a specific value, the corresponding Shapley value is the difference between the expected prediction (mean prediction for all features varied) and the mean of all predictions with that feature fixed at the value mentioned before (partial dependence). In game-theoretic parlance, Shapley values, when added up, always provide the difference between the game outcome with all players present and with no players present (the difference between the actual prediction and the mean prediction). This is demonstrated in the top chart (\textit{waterfall plot}) of Fig. \ref{fig:shap}. The point prediction considered is one of the highest likelihood points in our obtained NS dataset. The lower panel of Fig. \ref{fig:shap} shows the distributions of all the Shapley values for each parameter with their relative impact on the $x$-axis. The feature values are shown with a gradient colour code.

These plots show that while the sensitivity of $\lambda_2$ and $\lambda_3$ varies considerably over their possible values, they have a large impact at the highest likelihood region. The impact of $\lambda$  varies less and it is mostly on the larger side. On the other hand, the analysis is not that sensitive to changes in $v_T$, $\lambda_1$, and $\lambda_4$ on the whole, especially at the \textit{best-fit} region.

\section{Summary and Conclusion}\label{sec:concl}
Sampling a high-dimensional parameter space, as is often encountered with various BSM scenarios, is a daunting task. The most widely used statistical tool to perform this and draw Bayesian inference is MCMC. Despite its effectiveness in performing the task, it remains computationally expensive to find all the modes when dealing with a multi-modal parameter space. Parallelization and automatic tuning of the convergence are also non-trivial. NS is a more efficient technique that can be used to alleviate these issues. NS creates a weighted sample for the posterior starting from the whole constrained prior and is better suited to handle multi-modal problems. Its convergence is automatically tunable and it is highly parallelizable. The algorithm, however, slows down increasingly as it approaches high-likelihood regions, thus increasing the time to converge. A rapidly decreasing volume of the \textit{live} region compared to the constrained prior is one of the main reasons for this slowdown. The main source of slowdown in the case of BSM searches, however, is the HEP spectrum generators. Though these need to be run at each parameter point, a large chunk of any BSM parameter space gives rise to un-physical particle spectra and is therefore discarded. To reduce unnecessary runs of the spectrum generator, we propose using an ML classifier trained to predict whether a random set of input parameters can give rise to a theoretically consistent particle spectrum. We use an ensemble SNN, obtained from a chosen structure after hyper-parameter tuning, for the classifier. In addition to this, we introduce a generative algorithm within the NS loop to iteratively learn the structure of the coupled space of parameters and the log-likelihood function. This produces a large number of high-likelihood candidates with increasing accuracy and calculating the actual likelihood on these prevents the slowdown as the algorithm approaches convergence.  
A special type of normalizing flow network, known as Real-NVP or RNVP, is used as the generative algorithm. This network is trained repetitively with {\it truth} data which contains the parameters and their actual likelihoods. This is trained not at each iteration, but only after the pooled dataset has enough samples saved. The convergence depends crucially on the way the training dataset is chosen from the pooled data. The method labelled {\it RNVP-SF} optimizes the runtime while generating the correct posterior. 

We have considered the Type-II Seesaw model as a representative BSM scenario. This is a well-motivated and widely studied BSM scenario from the viewpoint of neutrino mass generation, LFV and LNV decays, and a rich scalar sector. We only consider a relatively large triplet VEV ($10^{-4}$ - $3.5$ GeV) scenario that results in a suppression of LFV and LNV decays. Our focus has exclusively been on the posterior of scalar sector parameters, subjected to the 125 GeV Higgs data, $\rho$-parameter, and the oblique parameters. We present our results in terms of $2D$ marginal distributions on various parameter planes and show the impact of direct collider search results alongside them. We have also shown the preferred mass regions for the exotic scalars, allowed by all theoretical and experimental constraints. We observe that for all these allowed points up to $95\%$ CL, the exotic scalar masses lie above $\sim 350$ GeV with most of them clustered in the $\sim 400-800$ GeV range. The mass difference between the doubly and singly charged Higgs mass lies within $\lesssim 40$ GeV window on either side of zero. The mixing angle between the two CP-even Higgs for the $95\%$ allowed region of $v_T$ ($\sim 0.3-2.5$ GeV) is small (as expected) with $\sin\alpha\sim 10^{-3}$. We further check the updated collider constraints on doubly charged Higgs masses and observe that the most stringent constraint arises when they are pair-produced and decay into a pair of $W$-bosons. However, the favoured parameter region obtained here is yet to be probed at the existing luminosity. We also comment on how the favoured parameter space affects the choice of neutrino Yukawa matrix elements when confronted with the neutrino oscillation data with a zero and non-zero CP-violating phase. 

The sampling technique described here is only the first iteration of its kind and it has the potential to improve in various ways. On one hand, the generative algorithm can be improved with better networks (e.g. \textit{Masked Auto-regressive Flows} \cite{papamakarios2017masked}) in favour of more precise predictions; on the other hand, the simple region sampler can be upgraded (e.g. with multi-ellipsoid search algorithms, such as \textit{MultiNest} \cite{Feroz:2008xx}). Even in its infancy, the method is unique and extremely efficient, as is evident from the results. It was run in a single \textit{Windows} desktop, with SPheno called through a virtual UNIX environment (WSL). With a complete Linux implementation, it was found to be at least 5 times faster. This reduction in computational time and cost helps the user identify the model posterior easily, thus ensuring a precise prediction of the observables. As a result, the experimental collaborations can concentrate further on the favoured parameter space only rather than constructing simplified scenarios representing different parts of the parameter space. With the accumulation of more data, the parameter space can be restricted more effectively, resulting in even more precise predictions. Eventually, in the absence of any new physics signal, this provides an efficient way to discard a model completely. On the other hand, using existing data, one can use this framework to compare similar models and identify the most favoured candidate.

\section*{Acknowledgment}
RB, SM, and SKP would like to acknowledge support from ANRF (erstwhile DST-SERB), India (grant order no. CRG/2022/003208). SM would also like to acknowledge ANRF for grant order no. CRG/2023/008570. RB and SR acknowledge the invaluable insights and guidance gained during the \textit{Machine Learning for Particle and Astroparticle Physics (ML4HEP-2024)} Workshop, at the Institute of Physics, Bhubaneswar. The authors also want to thank Jyotishka Datta of Virginia Tech. for explaining to them the nuances of Nested Sampling.

%
\bibliographystyle{
h-physrev}
\bibliography{MLNST2SS}
%
%
%

\end{document}